\input{aipcheck}

\documentclass[
    ,final            % use final for the camera ready runs
  ]
  {aipproc}

\layoutstyle{8x11single}

\usepackage{graphicx}
\usepackage{amssymb}
\usepackage{amsmath}
\usepackage{relsize}

\def\ket#1{|#1\rangle}

\def\scal#1#2{\langle#1|#2\rangle}
\def\matr#1#2#3{\langle#1|#2|#3\rangle}
\def\ave#1{\langle #1\rangle}

\def\uvo#1{\lq\lq #1\rq\rq}
\def\jj1{j(j\!+\!1)}
\def\xx1#1#2{#1(#1\!#2\!1)}
\def\mpm1{m\!\pm\!1}
\begin{document}

\title{Understanding Chaos via Nuclei}

\classification{05.45.Ac, 05.45.Mt, 21.60.Ev}
\keywords      {Classical Chaos, Quantum Chaos, Collective Motions of Nuclei}

\author{Pavel Cejnar}{
  address={Institute of Particle and Nuclear Physics, Faculty of Mathematics and Physics, Charles University,\\ V Hole{\v s}ovi{\v c}k{\'a}ch 2, 18000  Prague,  Czech Republic}
}

\author{Pavel Str{\' a}nsk{\' y}}{
  address={Instituto de Ciencias Nucleares, Universidad Nacional Aut{\'o}noma de M{\'e}xico, 04510, M{\'e}xico, D.F., Mexico}
}

\begin{abstract}
We use two models of nuclear collective dynamics---the geometric collective model and the interacting boson model---to illustrate principles of classical and quantum chaos.
We propose these models as a suitable testing ground for further elaborations of the general theory of chaos in both classical and quantum domains.
\end{abstract}

\maketitle

%%%%%%%%%%%%%%%%%%%%%%%%%%%%%%%%%%%%%%%%%%%%
%% MAINMATTER
%%%%%%%%%%%%%%%%%%%%%%%%%%%%%%%%%%%%%%%%%%%%

%\subsection{<A subsection>}
%\subsubsection{<A subsubsection>}
%\paragraph{<A subsubsubsection>}

\section{Introduction}

Chaos in ordinary language means disorder, randomness, absence of law, and unpredictability.
It played the major role in philosophy of ancient Greeks.
Already Empedocles (cca.490--430 B.C.) viewed the real world, \uvo{cosmos}, as a combination of the world of perfect order, \uvo{sphairos} (today we may tentatively translate it as \uvo{symmetry}), and the world of complete disorder, \uvo{chaos}.
Since chaos seemed to be too elusive for any kind of rigorous description, the attention of science has for long been focused only to sphairos. 
This has changed over the past decades when numerous explorations have shown that chaos is not as lawless as originally thought.
At present, the term {\em Chaos Theory\/} belongs to the common speak \cite{cha}.

\uvo{Chaos Theory} in both classical and quantum physics \cite{Gut} is built on very sophisticated mathematics. 
It sticks on the original idea of randomness, but at the same time it creates a new type of universality in the representation of reality.
The universality of chaos is based on ergodic character of motions and implies high efficiency of the statistically oriented description.
Since majority of systems in nature are chaotic, at least to a considerable extent, this type of description is often the ultimate one.
The routes to more deterministic descriptions may exist but they are usually impassable.   
Chaos thus belongs to very fundamental subjects in physics.

Among the physical systems that materialize the signatures of chaos in nature, an important place is held by atomic nuclei.
Already the fact that even more than 80 years after the discovery of their composition we are not able to present a satisfactory microscopic theory of nuclear structure is a good reason to think that nuclei are indeed chaotic objects par excellence.\footnote
{
At least in the sense that quantitative predictions seem to be rather unstable with respect to variations of the underlying microscopic models.
}
It is not an accident that nuclei were at dawn of the field named quantum chaos.

However, atomic nuclei not only show well recognized signs of chaos, they also provide a variety of models that can be used to better understand the rules of chaos in a more general context.
Such a path is followed in this text.
Our aim is not to draw an exhaustive map of the overlap between physics of chaos and physics of nuclei---such reviews exist \cite{Wei09,Mit10} and we may only recommend to read them.
Instead, we want to make a passage through selected topics of classical and quantum chaos using models taken from the description of nuclear collective dynamics.

\section{Nuclear Collective Hamiltonians}

Before diving into the ocean of chaos, we have to briefly introduce the models we will be playing with. 
These are the old (and the good) Geometric Collective Model (below abbreviated as GCM) and the newer and more sophisticated Interacting Boson Model (IBM) of nuclei.
Both these models attempt to describe nuclear collective motions---vibrations and rotations---assuming only the quadrupole type of nuclear deformations.
The number of degrees of freedom associated with both GCM and IBM is five.
Two degrees of freedom correspond to the quadrupole deformation of the nucleus in the body-fixed frame (one can think of an ellipsoid with a fixed volume); their dynamics therefore represents vibrational motions. 
The other three degrees of freedom correspond to the orientation of the whole nucleus in the laboratory frame---they describe rotations.
In the following we will mostly consider only the non-rotational case, i.e., fix the angular momentum (which is an integral of motions) to zero.
In this case, we will be left just with two vibrational degrees of freedom, opening a direct link to other $f=2$ systems famously studied in the context of chaos.
We will see that nuclear vibrations represent a rather rich specimen of chaos in two dimensions.   

The GCM, introduced already in 1952 by A.\,Bohr (for the review see \cite{Rowe}), treats the nucleus as structureless drop described by quadrupole shape variables $(\beta,\gamma)$ and rotation Euler angles $(\theta_1,\theta_2,\theta_3)$.
These 5 generalized coordinates are associated with a particular parametrization of the rank-2 tensor $\alpha^{(2)}$ (real, symmetric, and traceless in the Cartesian form), describing the quadrupole deformation of the nucleus in the laboratory frame.
The deformation tensor $\alpha^{(2)}$ can be introduced via the expansion of the nuclear radius $R(\vartheta,\varphi)$ to spherical harmonics $Y^{(\lambda)}_\mu(\vartheta,\varphi)$, or alternatively via  multipole moments of the nuclear mass (charge) distribution $\rho(\vec{r})$. 
In the first case, the $\alpha^{(2)}_\mu$ spherical component of the deformation tensor ($\mu=-2$,$\dots$,$+2$) is associated with the coefficient (up to complex conjugation) at the $Y^{(\lambda=2)}_\mu$ term of the radius expansion.
In the second case, the deformation tensor is obtained via the following integral (up to a scaling factor) of the mass (charge) density: $\alpha^{(\lambda)}_\mu\propto\int\rho(\vec{r})\,r^\lambda Y^{(\lambda)}_\mu(\vec{r}/r)\ d\vec{r}$.

Any variables describing the shape of the nucleus must be invariant under rotations of the nucleus as a whole.
There are just two independent building blocks for all rotational invariants made of the quadrupole deformation tensor---namely the quadratic and cubic couplings of the quadrupole tensor to zero total angular momentum.
The shape variables $\beta$ and $\gamma$ can be defined as a certain parameterization of these two elementary couplings:\footnote
{
We use the standard definition of the coupling $[a^{(\lambda_1)}\times b^{(\lambda_2)}]^{(\lambda)}$ of rank-$\lambda_1$ and -$\lambda_2$ spherical tensors  $a$ and $b$ to the resulting spherical tensor of rank $\lambda$ (the definition is analogous to the elementary angular momentum coupling). Scalar product of spherical tensors is defined for $\lambda_1=\lambda_2$ as $(a^{(\lambda)}\cdot b^{(\lambda)})\equiv(-)^\lambda\sqrt{2\lambda+1}[a^{(\lambda)}\times b^{(\lambda)}]^{(0)}$.
}
\begin{equation}
[\alpha^{(2)}\times\alpha^{(2)}]^{(0)}=\tfrac{1}{\sqrt{5}}\ \beta^2
\ ,\qquad
[[\alpha^{(2)}\times\alpha^{(2)}]^{(2)}\times\alpha^{(2)}]^{(0)}=-\sqrt{7}\ \beta^3\cos 3\gamma
\label{bega}
\,.
\end{equation}
Alternatively, $\beta$ and $\gamma$ can be derived from the eigenvalues of $\alpha^{(2)}$, which read as $(\alpha_1,\alpha_2,\alpha_3)=\sqrt{5/4\pi}\bigl(\beta\cos[\gamma-2\pi/3]$, $\beta\cos\gamma$, $\beta\cos[\gamma+2\pi/3]\bigr)$. 
These values are the only components of the deformation tensor in the intrinsic frame (IF) of the nucleus (the frame connected with the principal axes of the quadrupole deformation), where the deformation tensor is diagonal.
The transformation between the IF and laboratory frame (LF) is expressed in terms of the Euler angles $(\theta_1,\theta_2,\theta_3)$, describing the orientation of the deformed nucleus in LF.

%$\beta\cos\gamma\equiv x=\alpha^{(2)}_0|_{\rm IF}$ and $\beta\sin\gamma\equiv y=\sqrt{2}{\rm Re}\alpha^{(2)}_{\pm 2}|_{\rm IF}$ (the other spherical components of the quadrupole tensor are zero in the IF).

\begin{figure}
\includegraphics[height=.35\textheight]{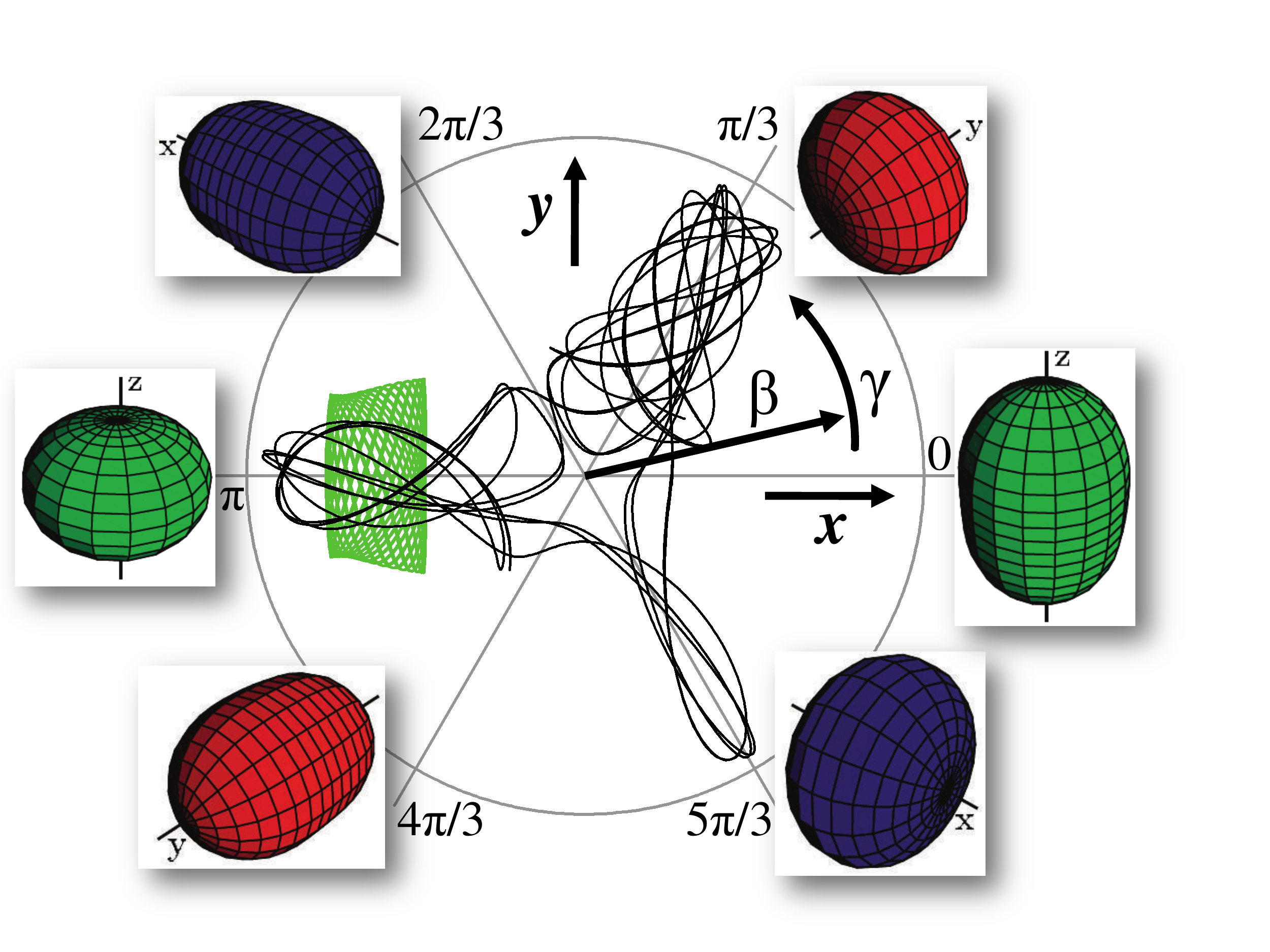}
\caption{The plane of Bohr deformation coordinates with a sample of associated shapes (only the shapes with an axial symmetry are drawn). 
Two examples of trajectories (chaotic and regular) represent some high-energy vibrations generated by the Hamiltonian (\ref{Hgcm}) with the same choice of parameters. 
Adapted from Ref.\,\cite{Cej11}.}
\label{begafi}
\end{figure}

Note that $(\beta,\gamma)$ can be considered as radial coordinates in the Cartesian plane $(x,y)\equiv(\beta\cos\gamma,\beta\sin\gamma)$.
While the radius $\beta$ quantifies the degree of deformation of the nucleus ($\beta=0$ for spherical shape), the angle $\gamma$ parameterizes the type of the deformed shape.
The plane is divided into six equivalent angular sectors, which follow from an inherent discrete symmetry of the problem, given by the dynamical equivalence of three possible orientations of the deformed shape in the IF (along $x,y,z$) and two deformation types (prolate, oblate).
The $x$, $y$, and $z$ axially symmetric prolate or oblates shapes are located at $\gamma$ equal to $(2\pi/3,5\pi/3)$, $(\pi/3,4\pi/3)$, and $(0,\pi)$, respectively, while the intermediate $\gamma$ values correspond to the related triaxial shapes.
Note that these conclusions are consistent with the above-given form of the deformation tensor in the intrinsic frame.
To become more familiar with the formalism (which, as one has to agree, looks rather tough as the first sight), the reader is encouraged to consult specialized literature, e.g., Ref.\,\cite{Rowe}.   
The plane of deformation coordinates with the basic sample of shapes is depicted in Fig.\,\ref{begafi}.
 
The momenta associated with the coordinates $\beta$ and $\gamma$ are denoted as $p_\beta$ and $p_\gamma$, respectively. 
In the non-rotational case, as noted above, the Euler angles $(\theta_1,\theta_2,\theta_3)$ are fixed, rendering zero values of the associated rotational momenta.
The resulting vibrational Hamiltonian therefore contains just the shape variables $(\beta,\gamma)$ and the corresponding momenta $(p_\beta,p_\gamma)$.
Our analysis is performed mostly with the Hamiltonian which contains purely quadratic momentum term of the kinetic energy, and quadratic, cubic and quartic coordinate terms of the potential energy:
\begin{equation}
H_{\rm vib}=\underbrace{\frac{1}{2M}\left[p_\beta^2+\left(\frac{p_\gamma}{\beta}^2\right)\right]}_{T_{\rm vib}}+\underbrace{A\beta^2+B\beta^3\cos 3\gamma+C\beta^4}_{V_{\rm vib}}\,.
\label{Hgcm}
\end{equation}
$A,B,C$ and $M$ are adjustable parameters satisfying physical constraints $M>0$ (kinetic energy is positive) and $C>0$ (the potential is confining).

The geometric Hamiltonian (\ref{Hgcm}) is closely related to the famous H{\'e}non-Heiles Hamiltonian, which was originally introduced in the context of stellar dynamics in galaxies and became a paradigmatic example of chaos in the classical domain \cite{HeHe}.
The  H{\'e}non-Heiles Hamiltonian is a special case of Eq.\,(\ref{Hgcm}) with $C=0$.
In this contribution (as in Ref.\,\cite{Cej04}) we claim that the inclusion of the confining $\beta^4$ term generates even much richer structures than those obtained in the $C=0$ case.
The coexistence of regular and chaotic types of classical motions generated by the full geometric Hamiltonian at high energies  is illustrated in Fig.\,\ref{begafi}.
The two trajectories shown in this figure were calculated for the same model parameters and energy, but represent very dissimilar species of collective motions.
The regular orbit corresponds to a moderate, highly organized vibration, localized in a relatively small region around an axially symmetric equilibrium shape (above the potential well to which the nucleus with given parameters drops in the low energy limit).
In contrast, the chaotic orbit represents some large-amplitude vibrations, which rather disorderly wander through small, medium and large oblate deformations with all three intrinsic orientations.

A generalization of the simple Hamiltonian (\ref{Hgcm}) can be achieved by considering higher-order terms in kinetic or potential energy.
We consider two extensions of the kinetic energy, particularly \cite{PStr}:
\begin{equation}
T_{\rm vib}^{(I)}=\frac{1+\kappa\beta^2}{2M}\left[p_\beta^2+\left(\frac{p_\gamma}{\beta}\right)^2\right]
\,,\qquad
T_{\rm vib}^{(II)}=\frac{1}{2M(1+\kappa\beta^2)}\left[p_\beta^2+\left(\frac{p_\gamma}{\beta}\right)^2\right]
\,,
\label{Halt}
\end{equation}
both based on some specific deformation-dependent effective mass ($\kappa$ is an additional model parameter). 
Such or similar extensions of the effective mass are relevant for the description of nuclear rotational bands since they modify the behavior of nuclear moments of inertia.
Here we consider them to probe the sensitivity of the vibrational measures of chaos to the general form of the Hamiltonian kinetic term.

Another extension of the above Hamiltonian is the inclusion of rotational terms, containing the intrinsic components of angular momentum and the corresponding moments of inertia. 
This however leads beyond the 2D description and makes the analysis much more difficult. 
We follow this line only in rather a restricted manner---considering only rotations around a fixed axis $z$ \cite{Str06}.
This can be done by introducing a third collective coordinate, angle $\delta$, and the associate momentum $p_\delta$, which coincides with the angular momentum of the nucleus around $z$ in both laboratory and intrinsic frames.
The quadrupole deformation tensor $\alpha^{(2)}|_{\rm LF}$ in the laboratory frame is obtained from the diagonal form $\alpha^{(2)}|_{\rm IF}$ in the intrinsic frame (see above) via rotation by angle $\theta_3=\delta/2$ around the $z$-axis.
%The coordinate triple  $(\beta\cos\gamma,\beta\sin\gamma\cos\delta,\beta\sin\gamma\sin\delta)\equiv(x,y,z)$ now represents three nonzero intrinsic elements of the quadrupole tensor 
%$(\alpha^{(2)}_0,\sqrt{2}{\rm Re}\alpha^{(2)}_{\pm 2},\sqrt{2}{\rm Im}\alpha^{(2)}_{\pm 2})|_{\rm IF}$, but eigenvalues of the new tensor still depend only on the deformation variables $\beta$ and $\gamma$. 
This identifies $\delta$ as a variable connected with the IF $\leftrightarrow$ LF transition.
The rotational energy reads as: 
\begin{equation}
T_{\rm rot}=\frac{1}{2M\beta^2\sin^2\gamma}\ p_\delta^2\,,
\label{Trot}
\end{equation}
which has to be added to the vibrational Hamiltonian in Eq.\,(\ref{Hgcm}) to get the total Hamiltonian $H=T_{\rm rot}+T_{\rm vib}+V_{\rm vib}$.
Details can be found in Ref.\,\cite{Str06}.

An important feature of the geometric Hamiltonian is related to its scaling properties---the fact that dynamics of the system must be invariant under scale transformations affecting the units of relevant quantities, i.e., energy, coordinate, and momentum.
Let us return now to the bare vibrational Hamiltonian (\ref{Hgcm}).
Since the scale transformations of this form can be absorbed in parameters $(M,A,B,C)$, we deduce that some choices of these parameters must be dynamically equivalent.
A more detailed treatment \cite{Str06} shows that the quantity $\tau=B^2/(AC)$ is invariant under all scale transformation and represents the only essential parameter of the bare vibrational model in the classical case.
The value of the mass parameter $M$ is relevant only in the quantum case, weighting the absolute density of quantum states in the discrete energy spectrum (for $M\to\infty$ we get the classical limit with infinite state density).
This effective parameter reduction implies a crucial simplification of the numerical analysis of the geometric model.

The quantization of the nuclear geometric model is achieved by replacing individual terms in the GCM Hamiltonian by the corresponding operators in the Hilbert space of wavefunctions $\Psi(\beta,\gamma,\theta_1,\theta_2,\theta_3)$.
This task is mathematically rather involved and was completed with a substantial contribution of the patron of this School, Marcos Moshinsky---see, e.g., Refs.\,\cite{Cha76,Mos77} and many others.
The basic form of the vibrational kinetic energy reads as follows:
\begin{equation}
\hat{T}_{\rm vib}=-\frac{\hbar^2}{2M}\left(\frac{1}{\beta^4}\frac{\partial}{\partial\beta}\beta^4\frac{\partial}{\partial\beta}+\frac{1}{\beta^2\sin3\gamma}\frac{\partial}{\partial\gamma}\sin 3\gamma\frac{\partial}{\partial\gamma}\right)
\label{TQgcm}
\,.
\end{equation}
Note that this operator acts in the space where the scalar product $\scal{\Psi_1}{\Psi_1}$ is given by an integration over the collective coordinates with a specific measure \cite{Mos77}.
The expression of the potential $\hat{V}_{\rm vib}\equiv V_{\rm vib}(\beta,\gamma)$ is the same as in the classical case, see Eq.\,(\ref{Hgcm}).
Since we are mostly interested in the non-rotational motions, the rotational energy will not be included into the quantized Hamiltonian.

Examining Eq.\,(\ref{TQgcm}) one immediately notices the differences from the kinetic energy of a particle in a 2D space, which in polar coordinates $(r,\vartheta)$ reads as 
%$\hat{T}=-\frac{\hbar^2}{2M}\left(\frac{1}{r}\frac{\partial}{\partial r}r\frac{\partial}{\partial r}+\frac{1}{r^2}\frac{\partial^2}{\partial\vartheta^2}\right)$.
$\hat{T}\propto r^{-1}(\partial/\partial r)r(\partial/\partial r)+r^{-2}(\partial^2/\partial\vartheta^2)$.
The absence of Euler angles in the description of vibrational motions suggests a possibility to quantize the vibrational energy in the standard 2D manner, i.e., using the formula:
\begin{equation}
\hat{T}_{\rm vib}=-\frac{\hbar^2}{2M}\left(\frac{1}{\beta}\frac{\partial}{\partial\beta}\beta\frac{\partial}{\partial\beta}+\frac{1}{\beta^2}\frac{\partial^2}{\partial\gamma^2}\right)
\label{TQgcm2}
\,.
\end{equation}
Note that this implies a change of the definition of scalar product and the necessity to impose the condition $\Psi(\beta,\gamma-2\pi/3)=\Psi(\beta,\gamma)=\Psi(\beta,\gamma+2\pi/3)$ to keep link with the 5D quantization, where this condition is naturally satisfied.
In addition, we choose one of two possible angle-reflection symmetries of the wavefunctions, namely $\Psi(\beta,\gamma)=\pm\Psi(\beta,\gamma)$.
We therefore construct three different quantizations of the vibrational collective Hamiltonian, all of them yielding the same classical limit: the 2D-even, 2D-odd, and the 5D quantizations (in the 5D case, the reflection symmetry is fixed).
From the viewpoint of nuclear theory, only the 5D option is correct, but from the viewpoint of chaos theory (which is in our focus here) we have the freedom to probe all three possibilities.
We will use this freedom to check the influence of the quantization scheme on the signatures of quantum chaos. 
For details, see Ref.\,\cite{Str09a,Str09b}.

Let us move to the alternative description of nuclear collective motions with the aid of the interacting boson model.
The IBM, introduced in 1970's by F.\,Iachello and A.\,Arima (for a review see \cite{Iach}), does a similar job as the GCM, but in terms of an ensemble of bosons with angular momentum 0 ($s$-bosons) and 2 ($d$-bosons).
These bosons are commonly interpreted as an approximation for nucleon pairs in an even-even nucleus (their total number $N$ is usually fixed at a half of the number of valence nucleons), but at the same time they give rise to quadrupole degrees of freedom needed to describe basic collective motions of nuclei.
The $s$- and $d$-bosons live in a Hilbert space associated solely with their spin degree of freedom (such space has a finite dimension) and interact with each other through two-body interactions.
For instance, a popular form of the Hamiltonian is composed of the $d$-boson number operator $\hat{n}_d$ and a bosonic quadrupole operator $\hat{Q}_\chi$ as follows:
\begin{equation}
\hat{H}=\frac{A}{N}\underbrace{(d^\dag\cdot\tilde{d})}_{\hat{n}_d}
-\frac{B}{N(N-1)}\underbrace{(d^\dag s+s^\dag\tilde{d}+\chi[d^\dag\times\tilde{d}])^{(2)}}_{\hat{Q}_\chi}\cdot\underbrace{(d^\dag s+s^\dag\tilde{d}+\chi[d^\dag\times\tilde{d}])^{(2)}}_{\hat{Q}_\chi}
\,,
\label{Hibm}
\end{equation}
where we used $d$-boson annihilation operators $\tilde{d}_m=(-1)^m d_{-m}$ having proper transformation properties of a rank-2 tensor.
The parameters $A$ and $B$, weighting both terms in Eq.\,(\ref{Hibm}), are more or less arbitrary, while $\chi$ from the quadrupole operator varies within interval $[-\sqrt{7}/2,0)$ for prolate shapes and $(0,+\sqrt{7}/2]$ for oblate shapes.
Note that the $\hat{n}_d$ term of the above Hamiltonian is of the 1-body nature and the term $(\hat{Q}_\chi\cdot\hat{Q}_\chi)$ generates specific 2-body interactions. 

The link between the bosonic Hamiltonian {\`a} la (\ref{Hibm}) and the geometric Hamiltonian of the type (\ref{Hgcm}) goes via the use of coherent states in the IBM.
For instance, the Glauber type of coherent states with an average total number of boson equal to $N$ is given by:\footnote
{
Alternatively, one may use a projection of the coherent state (\ref{Glau}) onto a subspace with a sharp boson number, yielding a bosonic condensate state.  
}
\begin{equation}
\ket{a}={\cal N}_{a}\exp\left[\biggl(N-\sum_{\mu=-2}^{+2}|a_\mu|^2\biggr)^{\frac{1}{2}}s^\dag+\sum_{\mu=-2}^{+2}a_\mu d^\dag_\mu\right]\ket{0}
\label{Glau}
\,,
\end{equation}
where ${\cal N}_{a}$ stands for a normalization factor to ensure $\scal{a}{a}=1$.
The complex variables $a_\mu$ (with $\mu=-2,\dots,+2$) which parameterize the coherent state, have transformation properties of a rank-2 spherical tensor and can be associated with a linear combination of the deformation tensor $\alpha^{(2)}$ and the tensor of associated momenta $\pi^{(2)}$.
Performing the same gymnastics as in the GCM case (for details see Ref.\,\cite{HL82}), one obtains the five generalizad collective coordinates $(\beta,\gamma,\theta_1,\theta_2,\theta_3)$ and the corresponding momenta.
The general bosonic Hamiltonian $\hat{H}$ is then translated into a function of these canonical variables by evaluating its expectation value in the respective coherent state $H(a)\equiv\matr{a}{\hat{H}}{a}$.
In the $N\to\infty$ limit, this function coincides with the classical Hamiltonian $H_{\rm cl}$ associated with $\hat{H}$.

\begin{figure}
\includegraphics[height=0.45\textheight]{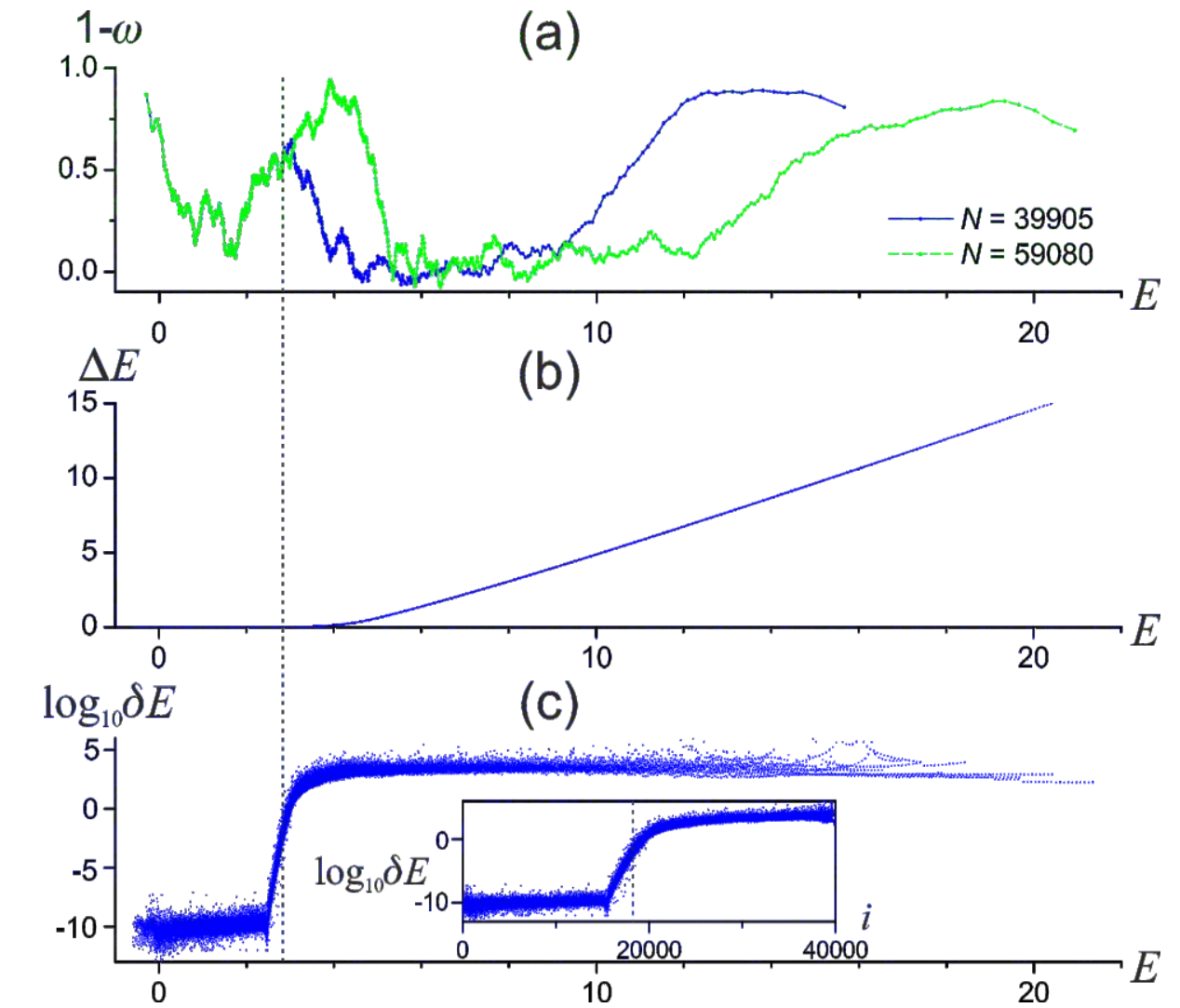} 
 \caption{Convergence properties of quantum GCM calculations. Comparison of results obtained for $(A,B,C)=(-1,0.62,1)$ and $\hbar^2/2M=1.25\cdot 10^{-5}$, using two truncation dimensions of the Hamltonian matrix expressed in the 2D oscillator basis, namely $N=$39905 and 59080. Panel (a) shows a quantum measure of chaos $1-\omega$ (where $\omega$ is so-called Brody parameter of the nearest-neighbor spacing distribution;  for explanation see below) calculated for spectra determined by a numerical diagonalization in both truncated bases. Panels (b) and (c) show the absolute and relative differences between individual level energies, $\Delta E$ and $\delta E=\Delta E/E$, respectively. It is seen that at a certain energy the two calculations yield diverging results. Only the fraction of the spectrum below the vertical dashed line (corresponding to less than a half of the smaller dimension) is reproduced correctly. Adapted from Ref.\cite{PStr}.}
\label{Egcm}
\end{figure}

For instance, Hamiltonian (\ref{Hibm}) yields the following classical limit \cite{Mac07}:
\begin{eqnarray}
H_{\rm cl}=
&&\!\!\!\!\!\!\!\!\!\!\!\!\!\!\!
A\frac{T+\beta^2}{2}-2B\beta^2\left(1-\frac{T+\beta^2}{2}\right)
-\frac{2\chi B}{\sqrt{7}}\left[\left(\frac{p_\gamma^2}{\beta}-\beta p_\beta^2-\beta^3\right)\cos3\gamma+2p_\beta p_\gamma\sin3\gamma\right]\sqrt{1-\frac{T+\beta^2}{2}}
\nonumber\\
&&\!\!\!\!\!\!\!\!\!\!\!\!\!\!\!
-\frac{2\chi^2 B}{7}\left[\left(\frac{T+\beta^2}{2}\right)^2-p_\gamma^2\right]
\,,
\qquad\qquad\qquad\qquad\qquad\qquad
T=p_\beta^2+\left(\frac{p_\gamma}{\beta}\right)^2
\label{Hibmclas}
\,.
\end{eqnarray}
It needs to be stressed, with regard to the boundedness of the IBM energy spectrum, that the coordinates and momenta appearing in the coherent-state parameterization are restricted only to finite domains, namely: $\beta\in[0,\sqrt{2}]$, $p_\beta\in[0,\sqrt{2}]$, and $p_\gamma\in[0,1]$.
Thus the square root in Eq.\,(\ref{Hibmclas}) is always well defined.
We may conclude that the IBM classical Hamiltonians are closely related to the GCM  Hamiltonians, discussed above, though by far not identical with them.
The most important difference between both models lies in the kinetic energy terms, which have a much more complex form in the IBM case.
On the other hand, the potential energy terms, which can be extracted from the Hamiltonian by setting all momenta to zero, are rather similar and yield analogous dynamical features.

In conclusion of this section we say a few words on numerical computations.
Classical trajectories for both GCM and IBM are determined with the aid of the fourth-order Runge-Kutta integration of the classical equations of motions. 
Some details can be found in Ref.\,\cite{PStr}.
Quantum spectra are obtained through a numerical diagonalization of the respective quantum Hamiltonian.   
Calculations within the IBM (which has just a finite dimensional Hilbert space) are naturally done in the complete basis composed of states with different numbers $s$- and $d$-bosons, coupled to definite angular momenta.
The calculations in GCM (infinite dimensional space) has to be performed  in a truncated basis.
The results of these  must be checked for convergence.
Let us streess that due to the restriction to the subset of the quantum spectrum with angular momentum $J=0$, we perform all GCM calculations in two dimensions.
In general, only a certain fraction of the GCM states obtained by the numerical diagonalization within a truncated 2D basis can be employed in the analyses.
As illustrated in Fig.\,\ref{Egcm}, these converged states yield unambiguous and reliable results.

\section{Classical Chaos}

In physics, we like most to deal with integrable systems.
These are the systems which---roughly speaking---have a sufficiently large degree of symmetry to make the dynamical equations solvable.
More precisely \cite{Arnold}, the full integrability of a classical Hamiltonian system with $f$ degrees of freedom requires the following three conditions to be simultaneously satisfied:
(a) there exists $f$ integrals of motions $F_i$ ($i=1,\dots f$), represented as functions in the $2f$-dimensional phase space which have zero Poisson bracket with the Hamiltonian, $\{F_i,H\}=0\,\forall i$, (b) all these integrals are mutually \uvo{in involution}, i.e., have vanishing Poisson brackets with each other, $\{F_i,F_j\}=0$ $\forall i,j$ (in quantum mechanics, we would say: they are compatible), and (c) the $2f$-dimensional gradients $\nabla F_i$ are all linearly independent in every point of the phase space.
These conditions guarantee that the Hamiltonian (as well as any other conserved quantity) can be written as a function of the given integrals, $H=H(F_1,..F_f)$, and that there exists a canonical transformation $(q_i,p_i)\mapsto(\theta_i,I_i)$ to some new variables (so-called action-angle variables) in terms of which the solution of the Hamilton equations becomes rather trivial (the transformation itself being usually highly nontrivial).

For example, the dynamics of the geometric collective model and the interacting boson model is integrable if the corresponding Hamiltonian does not depend on variable $\gamma$.
Then the momentum $p_\gamma$ becomes an integral of motion, in addition to the Hamiltonian $H$ itself.
It is easy to verify that all three conditions for integrability in dimension $f=2$ are fulfilled in this situation: (a) and (b) are obvious, and (c) follows from the fact that the gradient of $H$ is a nonvanishing normal to the energy surface, while the gradient of $p_\gamma$ is a nonvanishing tangent---therefore, they always define two perpendicular directions.

As we see from Eqs.\,(\ref{Hgcm}) and (\ref{Hibmclas}), the GCM and IBM Hamiltonians are independent of $\gamma$ (hence integrable) for $B=0$ and $\chi=0$, respectively.
The $p_\gamma$-dependent integral of motion is usually written in a less trivial form \cite{Cha76}:
\begin{equation}
\Lambda^2=p_\gamma^2+\sum_{k=1}^{3}\frac{L_k^2}{4\sin^2\left(\gamma-\frac{2\pi k}{3}\right)}
\label{senior}
\end{equation}
(where $L_k$ are components of the angular momentum $\vec{J}$ in the IF), which defines a quantity in the nuclear context called \uvo{O(5) seniority}.
Note that the rigorous proof of integrability in the full dimension $f=5$ (including rotational degrees of freedom) requires a construction of the fifth integral of motions, which is additional to $H$, $\Lambda^2$, $J^2$, and $J_z$. 
The IBM Hamiltonian (\ref{Hibm}) has an extra realization of integrability, situated at the points $(A,\chi)=(0,\pm\sqrt{7}/2)$ of the parameter space ($B$ remains an arbitrary scaling parameter).
It can be shown that this case of integrability of the IBM Hamiltonian is connected with its underlying SU(3) dynamical symmetry \cite{Iach}.

\begin{figure}
\includegraphics[height=0.35\textheight]{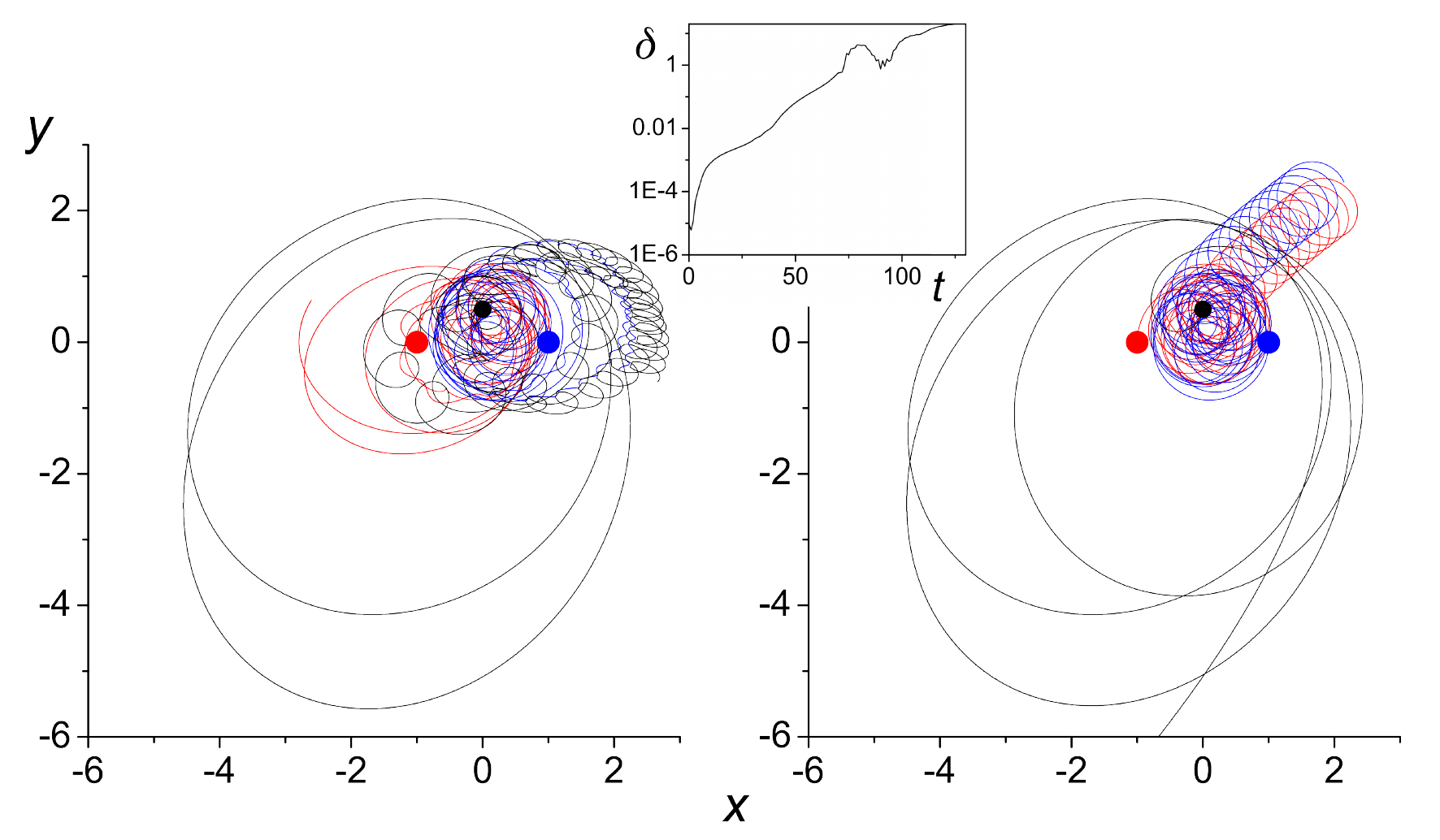}
\vspace{5cm}
  \caption{An illustration of chaos in the three body problem constrained to a plane. Left and right panels show two evolutions of the 3-body system with $m_1=m_2=1$ (red and blue dots) and $m_3=0.2$ (black dot). The two cases, differing just by a small displacement of the lighter body, show rather different sets of trajectories. Exponentially diverging distance between both alternative positions of the lighter body is shown in the inset. Adapted from Ref.\cite{PStr}.}
\label{body3}
\end{figure}

\begin{figure}
\includegraphics[height=0.3\textheight]{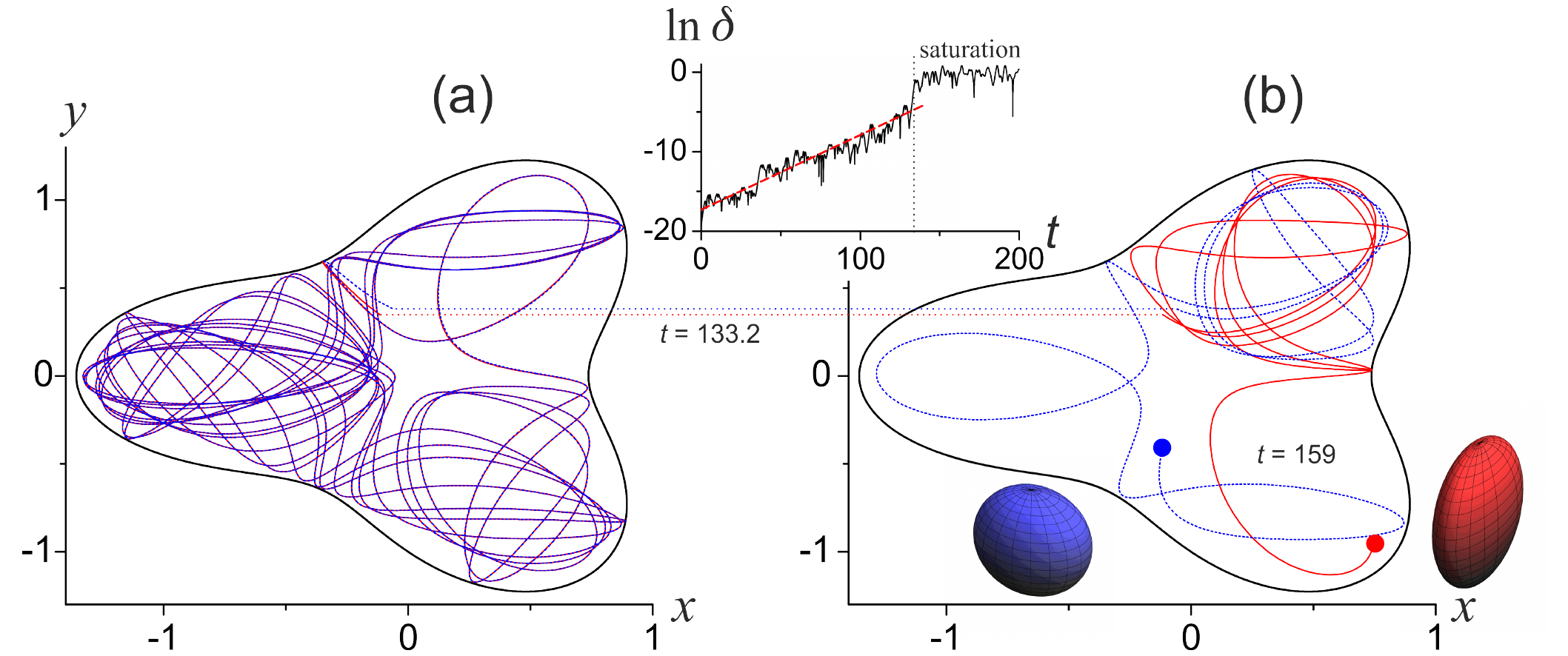}
\vspace{5cm}
  \caption{The instability of vibrational motions in the geometric model (analogous to Fig.\,\ref{body3} for 3-body system). Panel (a) shows evolution of initially close trajectories for times $t<133.2$ units, in which the deviation is not large. Panel (b) displays the evolution for times $133.2<t<159$ units, the two final positions represented by the blue and red dots (with the corresponding shapes drawn aside). Note that long-time deviation between the two vibrational orbits (given in the inset) reaches a saturation regime, which follows from the boundedness of the system for any finite energy.}
\label{Ginst}
\end{figure}

Already from these examples it must be obvious that integrability is a rather rare spice, therefore that by far not all the systems are integrable.
The full realization of this simple fact and its essential consequences came surprisingly late---only in the last decade of the 19th century.
It happened independently and almost simultaneously in the works of H.\,Poincar{\' e} and A.\,Lyapunov, who can undoubtedly be named the Founders of Chaos. 
Poincar{\' e}'s story, in particular, can be used as an exemplary illustration of the random ways in which science makes its progress \cite{Poincare}.
Poincar{\' e}, who in the preceding years spent a lot of effort to solve the three body problem, submitted an account of his results to a scientific competition, organized in 1885 to celebrate the birthday anniversary of King Oscar II of Sweden.
His work won the competition, but an assistant of one of the famous mathematicians in the jury noticed some unclear points in the derivations.
The thing was even much more complex than Poincar{\' e} thought.
In 1890, he submitted a new version of his treatise, in which the mathematical notion of chaos---defined in the sense of instability of motions---was born.
In one of the later writings Poincar{\' e} explains: {\it\uvo{It may happen that small differences in the initial conditions produce great ones in the final phenomena.}}
An example of this behavior is given in Fig.\,\ref{body3}. 
The new treatise was published at Poincar{\' e} own expenses, exceeding the reward of the competition, which may serve as a warning to those who think that science must always be profitable.
Lyapunov came to similar conclusions (concerning the instability of motions) in 1892 in his doctoral thesis \cite{Lyapunov}.

The description of chaos remained for long a rather exotic domain of abstract mathematics.
In integrable systems, the transformation to action-angle variables $(\theta_i,I_i)$ ensures that motions in the phase-space, described by Hamilton equations $dI_i/dt=-\partial H/\partial\theta_i=0$ and $d\theta_i/dt=\partial H/\partial I_i\equiv\omega_i$,  are localized on manifolds, which are topologically equivalent to tori. 
These motions are therefore highly organized.
Poincar{\' e} himself developed a method how the tori can be visualized for systems with $f=2$.
The method is based on cutting the phase space by a planar section and marking passages of individual trajectories through the section---today we call these pictures Poincar{\'e} maps.\footnote
{
The limitation to $f=2$ systems results from the requirement that each point of the planar section is crossed by a single trajectory.
Indeed, the point in the section is characterized by 2 variables (usually a coordinate $q_i$ and the associated momentum $p_i$), another variable is fixed by the section constraint (often of the form $q_j=0$), and the fourth variable (e.g., the momentum $p_j$) can---for Hamiltonian systems---be computed from energy conservation.
Therefore, we have determined all 4 variables in the 4-dimensional phase space. 
For $f>2$ systems with no integrals of motions additional to $H$ the Poincar{\' e} map would have to be more than 2-dimensional, which is rather impractical.
}
But what happens when the integrability is destroyed?
Certainly, some of the tori must die, but what is the exact scenario and an underlying mechanism?

\begin{figure}
\includegraphics[height=0.24\textheight]{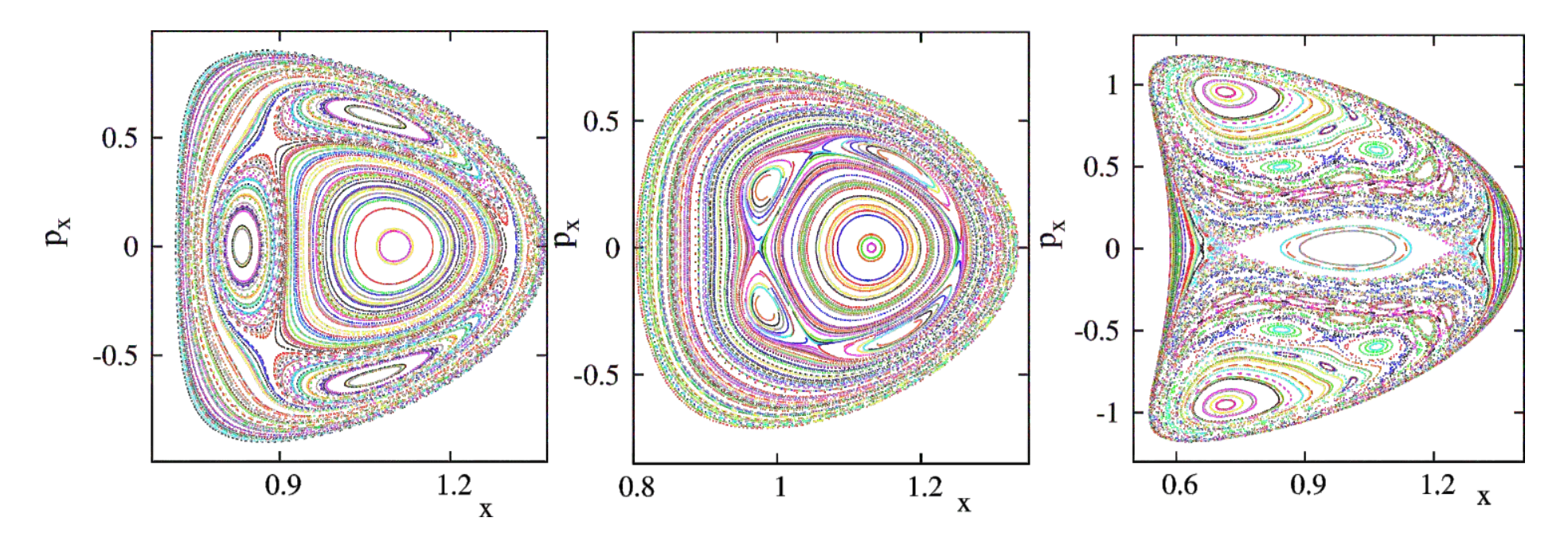} 
 \caption{Poincar{\' e} maps of the vibrational phase space of the interacting boson model showing various stages on the departure from integrability. 
The sections were obtained by calculating intersections of trajectories with the $y=0$ plane in the phase space; details of the Hamiltonian can be found in Ref.\,\cite{Mac12}.
The concentric curves correspond to deformed phase-space tori.
All panels show examples of the  Poincar{\' e}-Birkhoff fixed-point theorem.
Courtesy of M.\,Macek.}
\label{ibmsec}
\end{figure}

In 1912,  Poincar{\' e} conjectured a theorem, in which he partly attacked this problem.
The theorem was proven in 1913 and finalized in 1925 by G. Birkhoff and became known as the Poincar{\' e}-Birkhoff fixed-point theorem.
Consider an $f=2$ torus characterized by a pair of frequencies $(\omega_1,\omega_2)$ with ratio $\omega_1/\omega_2\equiv r$ .
If $r$ is an irrational number, the trajectory never closes a loop---the motion is only quasi-periodic.
In contrast, if $r$ is rational, all trajectories on the given torus are periodic---after $n$ crossings of the Poincar{\' e} surface they return to the initial point.   
The theorem of Poncar{\' e} and Birkhoff is focused on such a rational tori, or more precisely, on the remnants of these tori after the destruction of integrability by a general perturbation.
They showed that each \uvo{deceased} periodic torus leaves an even number of fixed points in the \uvo{$n$-fold} Poincar{\' e} map---a half of these fixed points is stable (elliptic), the other half is unstable (hyperbolic).
This was the first indication that the transition from the full regularity of an integrable system to a complete chaos of a strongly perturbed system is a highly non-trivial (and usually rather fascinating!) process.
Figure~\ref{ibmsec} shows some examples of Poncar{\' e} maps generated by the IBM  Hamiltonian with $J=0$ (hence $f=2$). 
Alternating islands of elliptic and hyperbolic fixed points are clearly visible.

Several decades later, an essential and famous result related to the order-to-chaos transition in classical Hamiltonian systems was formulated and proven by A.\,Kolmogorov (1954), V.\,Arnold (1963), and J.\,Moser (1962).
It is nowadays known as the KAM theorem \cite{Gut}.
What does it tell us?
It attacks the problem of what exactly happens if an integrable system is subject to an increasingly strong non-integrable perturbation.
Which of the tori of the original system die, for a given strength of perturbation, and which survive in a deformed form?
It turned out that the rational tori, such as those considered above, disappear immediately, leaving only the above described remnants.
On the other hand, some of the irrational tori are rather \uvo{tough guys}---they may survive (in a deformed form) even rather strong perturbations.
The clue for the torus to survive is that it must be \uvo{sufficiently irrational}.
This superficial statement can be quantified by means of the number theory, through the convergence properties of rational approximations of the number $r$ (the ratio of torus frequencies for an $f=2$ system).
The tori characterized by \uvo{more difficult} irrational ratios $r$ (slowly converging rational sequences) are more resistant than the tori with \uvo{easier} ratios (fast converging rational sequences).
Some people say that this law demonstrates the higher immunity of ugliness in the world, but sometimes even the most terrible irrational orbits look really pretty.

\begin{figure}
\includegraphics[height=0.35\textheight]{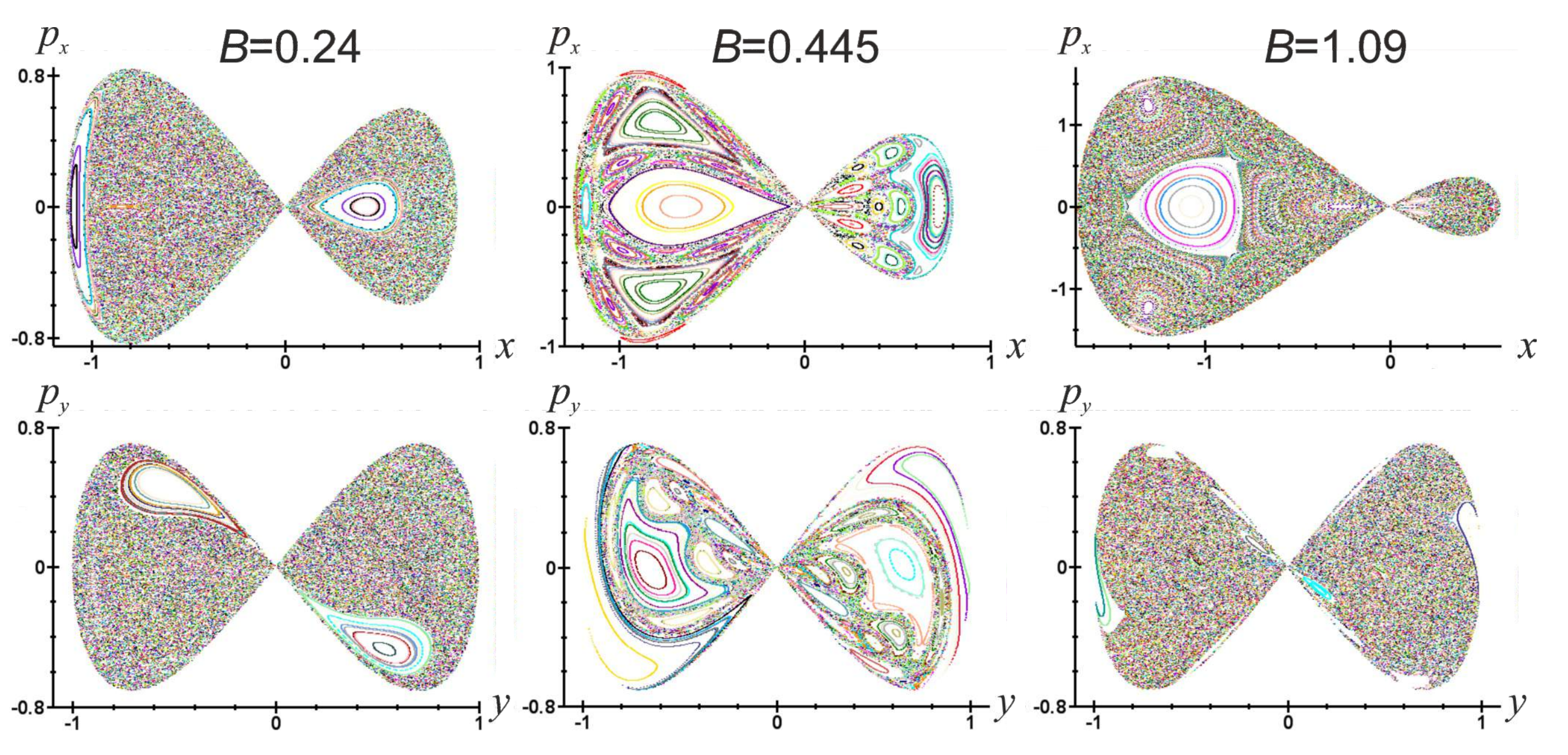} 
\caption{Poncar{\' e} maps of the vibrational phase space of the GCM Hamiltonian (\ref{Hgcm}) with $C=1=-A$. Value of $B$ is given in each column. The maps are constructed for energy $E=0$ of the potential local maximum at $\beta=0$ (here, all sections are contracted to a single point with $p_x=p_y=0$). The upper and lower rows show the $x=0$ and $y=0$ sections, respectively. Each map visualizes 1000 intersections of 100 trajectories with the plane of the section. Adopted from Ref.\cite{PStr}.}
\label{gcmsec}
\end{figure}

Several examples of complex Poncar{\' e} maps generated by the GCM vibrational Hamiltonian (\ref{Hgcm}) are shown in Fig.\,\ref{gcmsec}.
Particularly the maps in the middle column ($B=0.445$) show numerous fine structures, which must be caused by rather sophisticated regular orbits.
They appear in the system in spite of a large strength $B$ of the Hamiltonian term violating the full integrability at $B=0$.
However, a surprising observation is that most of these orbits cannot be identified with the survived tori from the $B=0$ system since the maps for lower values of $B$ (c.f., $B=0.24$) contain no trace of such orbits.
One may perhaps think of a \uvo{proximity} of the given Hamiltonian to an unknown integrable Hamiltonian beyond the parameter space of the GCM.
Note that also the $B=1.09$ maps contain some suspicious fine structures, whose origin cannot be explained via the only known integrable regime of the model.
Nuclear vibrations---even in their most simplistic description---are highly nontrivial type of motion!

To classify the overall degree of chaos for transitional systems such as those exemplified in Fig.\,\ref{gcmsec}, one needs to divide the whole phase space into the part filled with regular orbits, and the part filled with chaotic orbits.
Since each point of the phase space is crossed by a single trajectory, this division is unique, and one can introduce the degree of classical chaos through the following quantity, below shortened as the \uvo{regular fraction}:
\begin{equation}
f_{\rm reg}=\frac{\Omega_{\rm reg}}{\Omega_{\rm tot}}
\,,\qquad
\Omega_{\rm tot}=\int d^f\!p\ d^f\!x\ \delta[H(p,q)-E]
\label{freg}
\,.
\end{equation}
Here, $\Omega_{\rm tot}$ stands for the total \uvo{area} of the fixed-energy hypersurface $H=E$ in the $2f$-dimensional phase space, while $\Omega_{\rm reg}$ is the \uvo{area} occupied by regular orbits.
The fraction $f_{\rm reg}$ is therefore a quantity between 0 (for fully chaotic systems) and 1 (for completely regular systems).

\begin{figure}
\includegraphics[height=0.55\textheight]{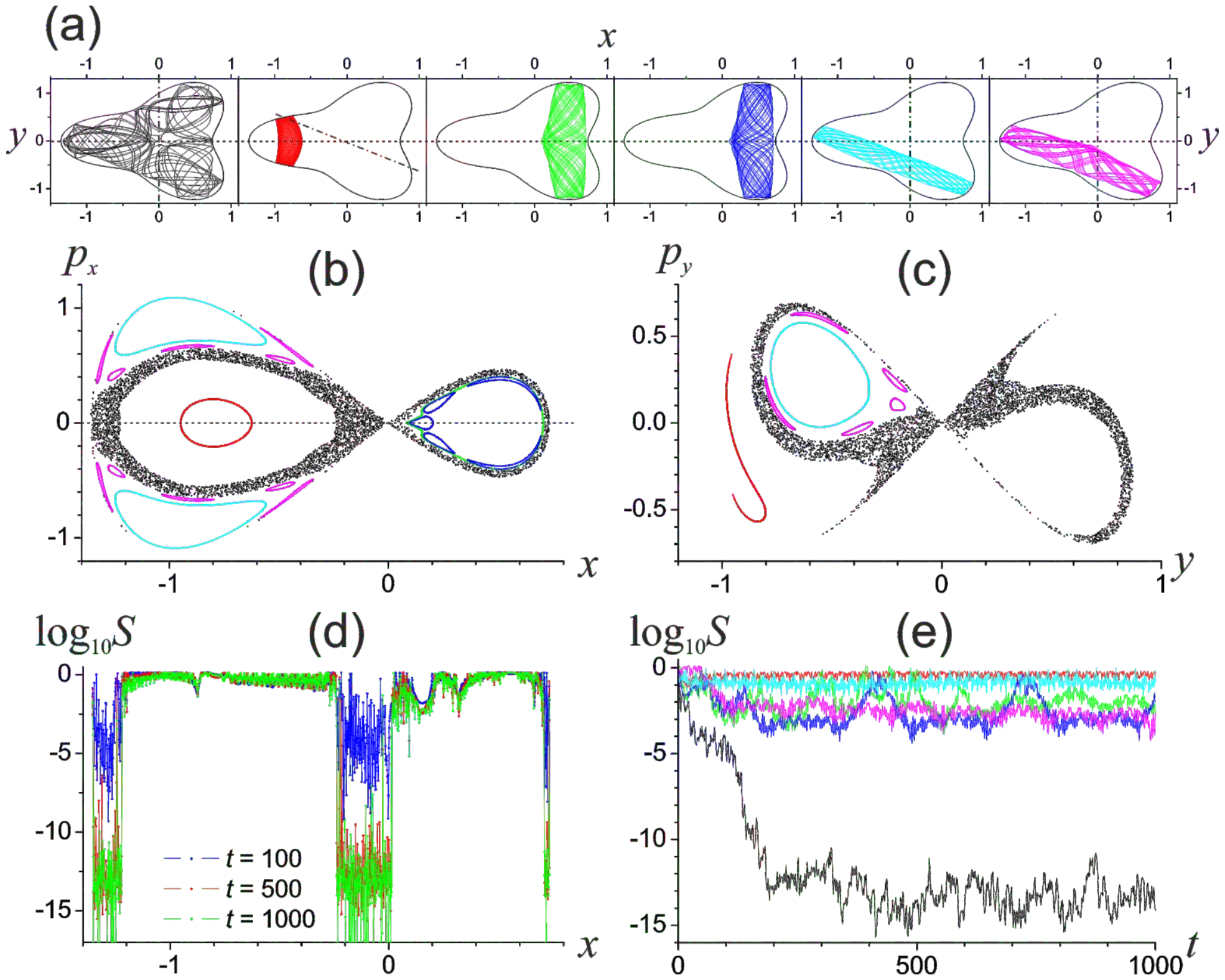} 
\caption{Efficiency of the alignment index technique in the GCM. Sample orbits are in panel (a) (the leftmost orbit chaotic, the others regular) and their Poincar{\'e} maps (for different section planes, as indicated in the upper panel) in panels (b) and (c).
The time dependence of the smaller alignment index $S$ corresponding to the given orbits is shown in panel (e), where we notice a clear difference between regular and chaotic cases. This is verified more systematically in panel (d), which shows $\log S$ at three different times for 2000 orbits initiating along the line $p_x=0$ of section (b). We observe that chaotic domains yield $S$ decreasing to negligible values, while in regular domains $S$ fluctuates around much higher values.   
Adapted from Ref.\cite{PStr}.}
\label{sali}
\end{figure}

It is clear that any numerical estimate of the regular fraction (\ref{freg}) for a system between order and chaos must be imperfect. 
Its precision depends primarily on the sampling density of the phase space by test trajectories.
Each of these trajectories is either regular (stable) or chaotic (unstable), and the regular fraction can be directly approximated by the fraction $f_{\rm reg}\approx N_{\rm reg}/N_{\rm tot}$.
The convergence of this estimate to the exact value of $f_{\rm reg}$ is guaranteed (assuming a uniform coverage of the phase space by initial points) in the \uvo{brute force} limit $N_{\rm tot}\to\infty$.
However, a truly hard computational problem is how to unmistakably determine the stability or instability of each individual orbit.
This requires to compute simultaneously the evolution of several orbits $X(t)\equiv[x(t),p(t)]$ (both $x$ and $p$ stand for $f$-dimensional vectors) which initiate in a close vicinity of the given reference orbit $X_0(t)\equiv[x_0(t),p_0(t)]$.
If the maximal Lyapunov exponent for the deviation $\Delta X(t)$ of these orbits is positive, the reference orbit is chaotic, if the Lyapunov exponent is zero, the orbit is regular.\footnote
{
The maximal Lyapunov exponent $\sigma$ is defined through the long-time phase-space distance $|\Delta X(t)|$ between a given orbit and an arbitrary accompanying orbit: 
$\sigma=\lim\limits_{t\to\infty}\max\limits_{\Delta X(0)}\frac{1}{t}\ln\frac{|\Delta X(t)|}{|\Delta X(0)|}$.
This yields an exponential growth of the maximal distance for long times: $\max\limits_{\Delta X(0)}|\Delta X(t)|\approx|\Delta X(0)|\exp(\sigma t)$.
Fof instance, the Lyapunov exponent for the trajectory in Fig.\,\ref{Ginst}, determined however only from a finite time interval of the linear growth of $\ln\delta$ (where $\delta$ stands for the distance in the coordinate plane), is $\sigma\approx 0.1$.
}
Applicability of this asymptotic-time criterion is however hindered in bounded systems, where the deviation of different orbits has an upper limit for any finite energy $E$, yielding therefore all Lyapunov exponents formally zero (cf. the time dependence of the deviation of the vibrational orbits in Fig.\,\ref{Ginst}).

Several  ways to bypass this computational problem has been proposed in the literature.
One of these ways---the one based on so-called alignment indices \cite{Sko01,Sko04}---was used in our GCM and IBM calculations.
The procedure resorts to evaluating only the relative phase-space deviations of trajectories instead of absolute ones: $\Delta(t)=\Delta X(t)/|\Delta X(t)|$, where $\Delta X(t)=X(t)-X_0(t)$.
The relative deviation is clearly a unity vector in the phase space, $|\Delta(t)|=1$, and only the mutual orientation of two or more such vectors is meaningful.
Let us take a pair of randomly selected orbits, both of them accompanying the reference orbit $X_0(t)$, and calculate relative deviations $\Delta_1(t)$ and $\Delta_2(t)$.
If the reference orbit is unstable, the relative deviations tend to align in either parallel or antiparallel directions---along the line of the maximal Lyapunov exponent.
This means that the smaller of the lengths $|\Delta_1(t)\pm\Delta_2(t)|$ (this quantity is called \uvo{smaller alignment index} and denoted below as $S$) converges to zero for long enough times.
In contrast, for a stable reference orbit $S$ keeps oscillating within the definition interval $[0,\sqrt{2}]$.
An illustration of the efficiency of this method in the GCM is presented in Fig.\,\ref{sali}.

\begin{figure}
\includegraphics[height=0.4\textheight]{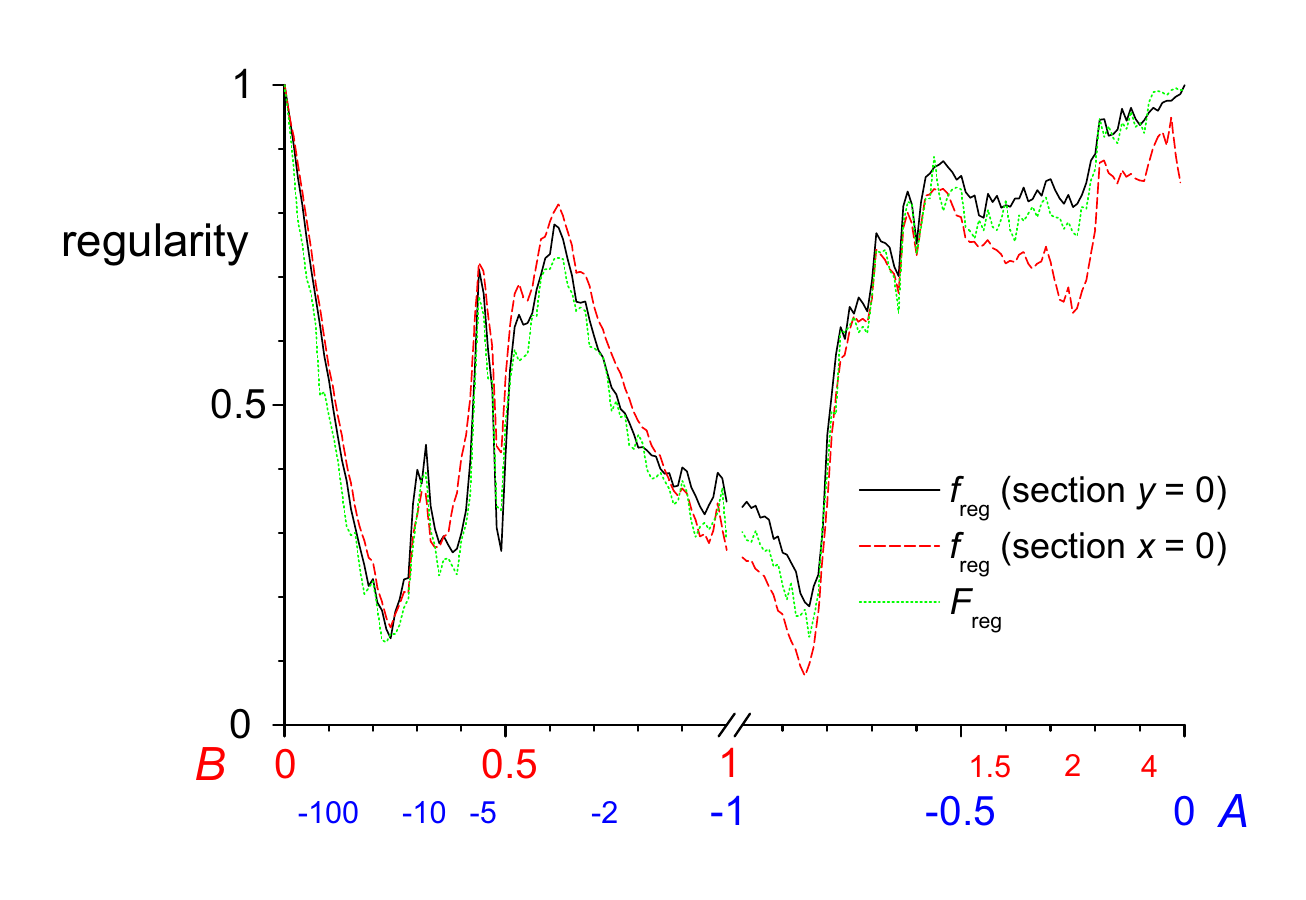} 
 \caption{The regular fraction $f_{\rm reg}$ for the GCM Hamiltonian (\ref{Hgcm}) at $E=0$ as a function of control parameter $A$ and/or $B$ for $C=1$. Three computation methods are compared by distinct curves. The methods are distingished by initiating the trajectories on the $x=0$ or $y=0$ sections (two curves denoted as $f_{\rm reg}$), or in the whole phase space (curve $F_{\rm reg}$). The horizontal axis is split into two parts: the first with $A=-1$ and $B\in[0,1]$ (the scale invariant parameter $\tau\in[0,-1]$), the second with $B=1$ and $A\in[-1,0]$ (hence $\tau\in[-1,-\infty]$). Adopted from Ref.\cite{PStr}.}
\label{freg0}
\end{figure}

Our procedure for calculating the regular fraction $f_{\rm reg}(E)$ for a given GCM or IBM  vibrational Hamiltonian ($f=2$) at fixed energy $E$ proceeds in the following steps:
(i) We choose either the $x=0$ or $y=0$ section of the phase space and calculate the domain in this section that can be reached by trajectories with the given energy.
(ii) We cover the accessible domain with tiny rectangular cells of the same area; total number of these cells is $N_{\rm tot}(E)$.
(iii) We perform a repeated computation in which an arbitrary point inside any cell of the accessible domain becomes an initial point of a calculated orbit (note that all 4 canonical variables are fixed by selecting a point in the section and energy).
The stability of this orbit is determined by the alignment index technique. 
If the orbit is stable, the initial cell as well as all the cells visited by the orbit during the evaluated time interval are counted and the resulting number is incremented to variable $N_{\rm reg}(E)$ (initially set to zero).
It can happen that a single cell is visited by both regular and chaotic orbits; in that case the value incremented to $N_{\rm reg}(E)$ for this cell is the properly weighted average between 0 and 1. 
(iv) The calculation is repeated until all cells of the accessible domain are either initial for a trajectory or are visited by at least one of these trajectories. 
(v) We estimate the regular fraction as $f_{\rm reg}(E)\approx N_{\rm reg}(E)/N_{\rm tot}(E)$.

This method has an advantage of a systematic coverage of the phase space---or more precisely, its 2D section---with trajectories.
On the other hand, it is clear that the resulting value of the regular fraction may depend on the phase-space section selected.
The efficiency of our procedure and its inherent ambiguities in the GCM are illustrated by Fig.\,\ref{freg0}, where we compare the values of $f_{\rm reg}(E)$ (for $E=0$ and variable control parameter $B$ and/or $A$) calculated for both $x=0$ and $y=0$ sections, as well as by directly counting the stable and unstable orbits generated without reference to any phase-space section.
It is seen that all three methods give approximately the same results.
The differences between the three curves in Fig.\,\ref{freg0} can be declared to be an internal uncertainty of the method at the selected level of resolution (number $N_{\rm tot}$).

Figure \ref{freg0}, apart from its methodological aspect, has also a very interesting physical content.
We see that for moderate values of parameter $B$ the regular fraction quickly drops from value $f_{\rm reg}=1$ at $B=0$ to $f_{\rm reg}\approx 0.1$ at $B\approx 0.2$.
In view of the KAM scenario, this is a perfectly expectable behavior since $B$ weights the nonintegrable perturbation in the Hamiltonian (\ref{Hgcm}).  
However, at $B\approx 0.25$ the regularity starts growing again!  
In particular, in the subsequent region up to $B\approx 0.7$ we observe a series of fine-structured peaks in the $f_{\rm reg}$ dependence.\footnote
{
The increase of regularity for $B\to\infty$ or $A\to 0$ is limited to energies $E\leq 0$ and results from a specific shape of the potential. 
}
The increased regularity in this parameter domain is connected with rather complex Poincar{\'e} maps, such as those shown  in the middle column of Fig.\,\ref{gcmsec}.
Quite surprisingly, the peaks in Fig.\,\ref{freg0} follow a certain \uvo{numerology} if expressed in scale invariant parameter $\tau$ (the four local peaks appear at integer multiples of the value $\tau$ corresponding to the first peak).    
The origin of this behavior is still unknown.
The complete dependence of $f_{\rm reg}$ on the GCM Hamiltonian parameters and energy can be found in Ref.\,\cite{Cej11}. 

\begin{figure}
\includegraphics[height=0.26\textheight]{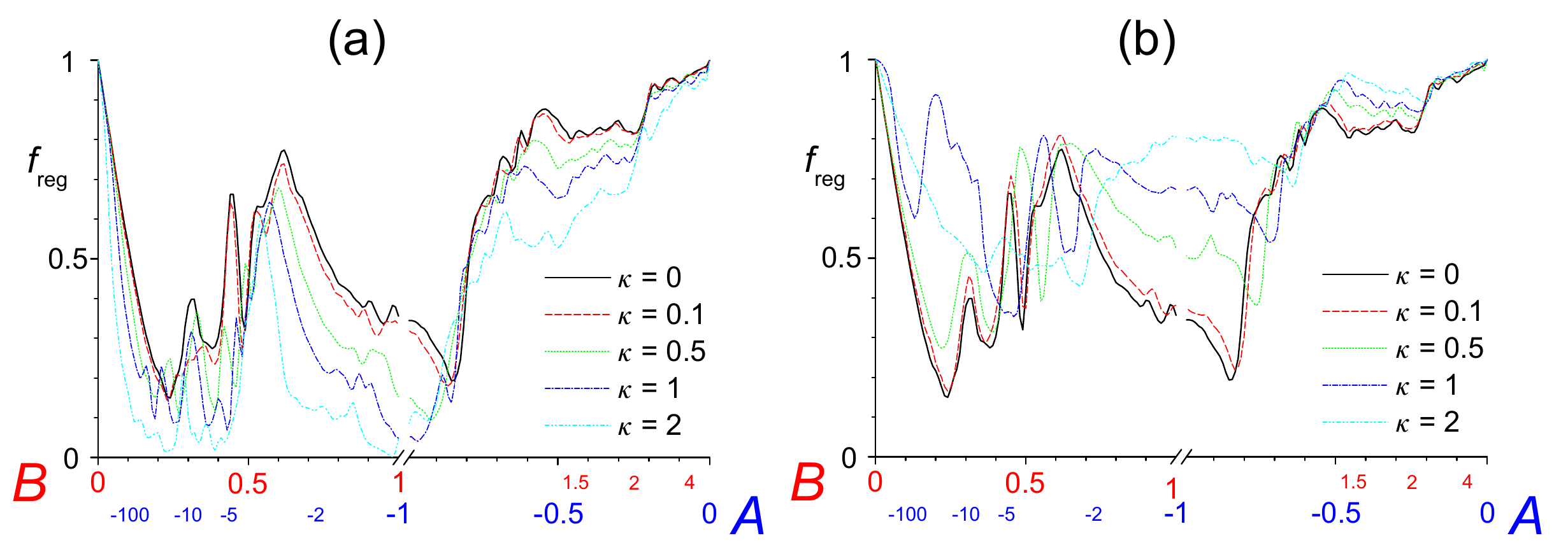} 
 \caption{The regular fraction at $E=0$ for GCM Hamiltonians with extended kinetic term $T_{\rm vib}^{(I)}$ (panel a) and $T_{\rm vib}^{(II)}$ (panel b), see Eq.\,(\ref{Halt}). Dependences for several values of parameter $\kappa$ are given for both types of kinetic term. The other parameters are the same as in Fig.\,\ref{freg0} (which corresponds to the $\kappa=0$ case in both panels). Adapted from Ref.\cite{PStr}.}
\label{freg0ext}
\end{figure}

In Fig.\,\ref{freg0ext}, we show the regularity plots for generalized GCM Hamiltonians with the kinetic energy terms according to Eq.\,(\ref{Halt}).
One can point out the opposite prevailing influence of type I and II extensions of the kinetic energy on the regular fraction: while the type I extension with growing $\kappa$ tends to reduce $f_{\rm reg}$,  the type II extension seems to rather increase it.
Note that the influence of the rotational term (\ref{Trot}) on the regularity plot was analyzed  in Ref.\,\cite{Str06}.

Let us note that similar structures as in Figs.\,\ref{freg0} and \ref{freg0ext} are observed also in the interacting boson model, where people often discuss about a so-called \uvo{arc of regularity} \cite{Mac07,Alh91,Alh93}. 
It needs to be stressed that the general IBM Hamiltonian is more sophisticated than the GCM one and therefore cannot be characterized by a single scale invariant parameter such as $\tau$.
In particular, the common parameterization (\ref{Hibm}) yields two essential parameters, namely $\chi$ and the proportion between $A$ and $B$, so that the peak series from  Fig.\,\ref{freg0} becomes a kind of \uvo{mountain ridge} in the regularity landscape.
The ridge follows an arc shape in the parameter plane and, as in the GCM, it is most pronounced for energies just above the $\beta=0$ local maximum of the potential.
The physics behind both phenomena is probably rather similar.

\section{Quantum chaos}

There is a fundamental problem to define chaos in quantum mechanics since linearity of the theory excludes any chance to introduce an exponential sensitivity to initial conditions.
Consider two normalized initial states $\ket{\Psi_0}$ and $\ket{\Psi'_0}$ differing by $\delta\ket{\Psi^{\bot}_0}$ with $\delta\ll 1$ and $\scal{\Psi^{\bot}_0}{\Psi^{\bot}_0}=1$, $\scal{\Psi_0}{\Psi^{\bot}_0}=0$.
The evolution of both states reads:
\begin{equation}
\begin{array}{rcl}
\ket{\Psi_0}&\to&\ket{\Psi_t}\equiv\exp(-i\hat{H}t/\hbar)\ket{\Psi_0}\,,\\
\sqrt{1-\delta^2}\ket{\Psi_0}+\delta\ket{\Psi^{\bot}_0}\equiv\ket{\Psi'_0}&\to&\ket{\Psi'_t}=\sqrt{1-\delta^2}\ket{\Psi_t}+\delta\ket{\Psi^{\bot}_t}
\,,
\end{array}
\label{Qevo}
\end{equation}
where all scalar products remain conserved due to unitarity of the evolution operator.
As a consequence, the squared distance of both states, $\Delta^2=||\Psi_t-\Psi'_t||^2\approx\delta^2$, is independent of time.
Quantum Hilbert space does not permit any kind of butterfly-wing effect.

A different type of instability, the one which remains relevant even in the quantum domain, was pointed out by A.\,Peres in 1984 \cite{Per84a}.
Instead of the dependence on initial conditions he proposed to consider the dependence of quantum dynamics on details of the Hamiltonian (tiny changes of interaction parameters or small external perturbations).
Although the resulting concept of so-called \uvo{fidelity} plays an important role in  many branches of quantum physics, including quantum information applications, it does not provide---unfortunately!---a selfsufficient definition of chaos in general quantum systems \cite{Pros}.
According to present understanding, chaos is firmly anchored in classical physics. 
Quantum physics can only explore a multitude of its distinct consequences (signatures) on the quantum level of description.  
Quantum chaos \uvo{defined} in this sense is therefore not a \uvo{phenomenon}, but rather a research field located somewhere in the highlands along the border between quantum and classical physics.
M.\,Berry, who undertook some pioneering expeditions to these wild territories, speaks about \uvo{quantum chaology} instead of quantum chaos \cite{Ber}.

For bounded quantum systems---those with discrete energy levels---the most important signatures of chaos were identified in statistical properties of energy spectra.
The relevant mathematics was created already in the 1950's, long before the quest for quantum chaos started \cite{rmt1}.
It was when E.\,Wigner looked for an achievable type of physical description related to long sequences of neutron resonances---plentiful and rather complex excited states of compound atomic nuclei.
Wigner decided to give up the exact description---prediction of the properties of all individual resonances---and proposed to strictly separate smooth (predictable) and fluctuating (statistical) parts of the corresponding quantities.
While the smoothened dependences (averages, which reflect only sort of \uvo{bulk} properties of the system) can be described in terms of some reasonable simplifying models, the fluctuations (that depend on all details of the dynamics) represents just random, unpredictable component of the problem.
In the 1980's it became clear that quantum signatures of chaos lie right in these fluctuating properties \cite{Boh84}. 

As an example, consider a discrete energy spectrum described by the state density $\rho(E)=\sum_n\delta(E-E_n)$.
To predict the position of every individual level $E_n$  might be too hard (or impossible---especially if the system is as complex as a nucleus.
However, one can perhaps estimate a smoothened density $\bar{\rho}(E)=\int dE'\,G_\sigma(E-E')\rho(E')$, where $G_\sigma(E\!-\!E')$ stands for a suitable smoothening function (e.g., the Gaussian of width $\sigma$).
Indeed, the smoothed density is closely related to the size $\Omega_{\rm tot}(E)$ of the energy shell in the multidimensional phase space in units associated with the Planck constant, which, in principle, can be evaluated with the aid of suitable techniques (like the Fermi gas approximation in the case of a nucleus).
In contrast, the remaining component of the state density, its fluctuating part $\tilde{\rho}(E)$, depends on tiny details of the Hamiltonian and cannot be predicted by simple models.\footnote
{
The fluctuating part of the state density can be given in terms of so-called trace formulae, which contain sums over classical periodic orbits of the system \cite{Gut}. It therefore depends on the most subtle details of classical dynamics. 
}
Nevertheless, statistical features of $\tilde{\rho}(E)$ reflect the degree of chaos in the classical limit of the system.

To extract the chaos-related information, one usually performs a transformation of the spectrum called \uvo{unfolding} \cite{rmt2}.
It can be written in the following way:
\begin{equation}
E_n\ \mapsto\ e_n=\int_{-\infty}^{E_n}dE'\ \bar{\rho}(E')
\quad\Rightarrow\quad
\left\{
\begin{array}{l}
\ave{e_n}=n\,,\\
\ave{e_{n+1}-e_n}=1\,.
\end{array}
\right.
\label{unfo}
\end{equation}
The unfolded sequence $\{e_n\}$ has a constant average density (equal to one) and therefore materializes statistical properties of the spectrum in the most explicit form.
It turns out that mutual correlations between the members of this sequence show some universal behavior, which depends on whether the classical counterpart of the system is regular or chaotic.
Surprisingly, in regular (integrable) systems the unfolded sequences of energies with the same quantum numbers (such as angular momentum or parity) exhibit only weak or virtually no correlations, making the visual appearance of spectra rather disordered. 
In contrast, chaotic systems yield strong correlations and produce seemingly more regular spectra.

\begin{figure}
\includegraphics[height=0.6\textheight]{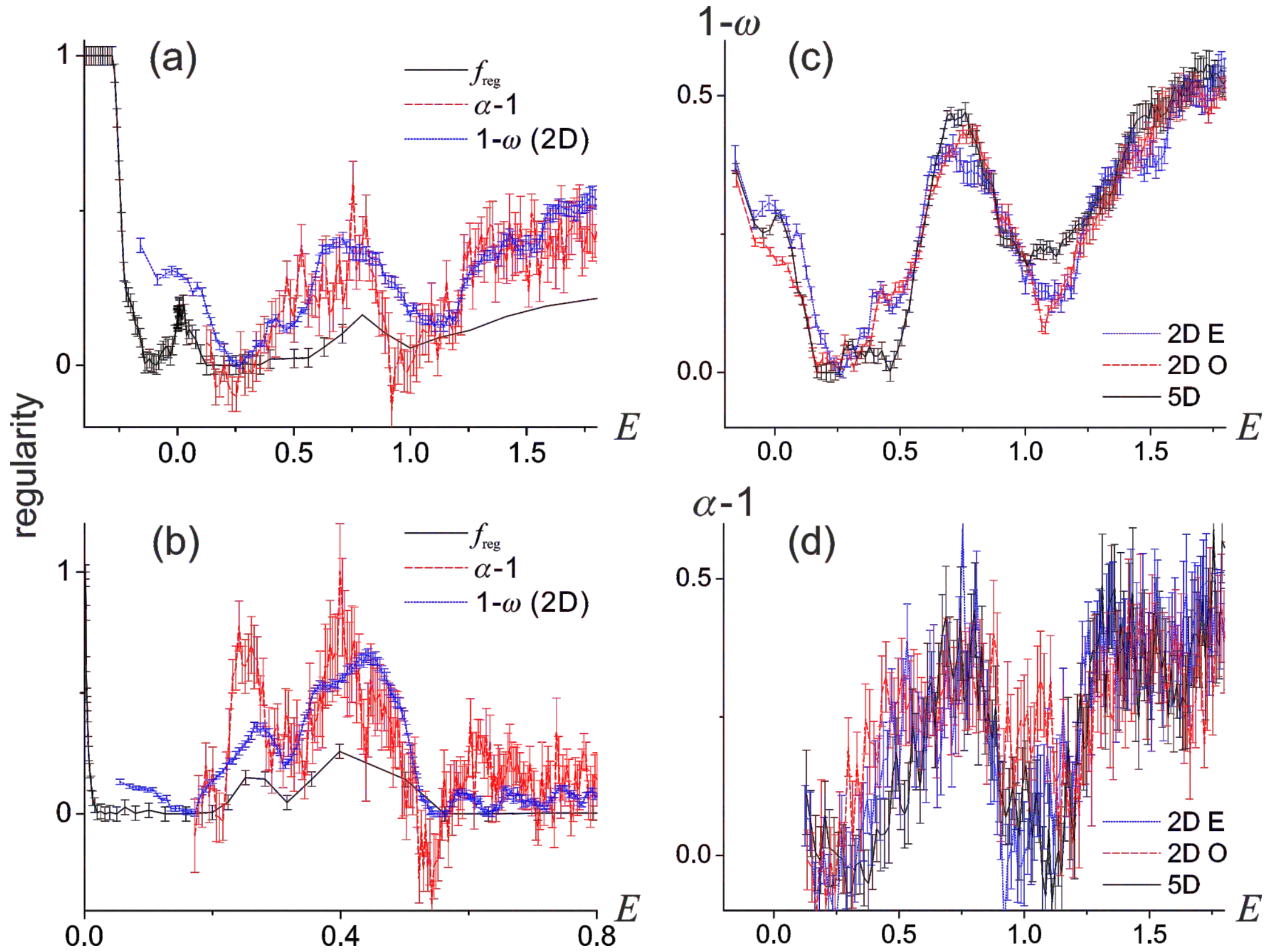}
\caption{Comparison of energy dependences of $f_{\rm reg}$ (classical regular fraction), $1-\omega$ (the Brody parameter adjunct), and $\alpha-1$ (a shifted exponent of the power spectrum) for the GCM. Panels (a) and (b) show results for $(A,B,C)=(-1,0.24,1)$ and for $(0.25,1,1)$, respectively. The mass parameter was chosen such that the displayed energy intervals contain 30 (panel~a) and 40 (panel~b) thousand of quantum levels, calculated with the 2D quantization option (see the text).  Panels (c) and (d) show separately $1-\omega$ and $\alpha-1$ for all three different quantization schemes (the Hamiltonian parameters are the same as in the respective left panel). We conclude that---within the error bars (given by the fitting methods of the respective quantities in finite samples of levels in a vicinity each energy)---the correlation measures show no dependence on the quantization.}
%\vspace{-2cm}
\label{Qchagcm}
\end{figure}

The correlations in unfolded spectra of chaotic systems are of both short- and long-range types.
The short range correlations are usually expressed through the statistical distribution of nearest neighbor spacing $s=e_{n+1}-e_n$ constructed for sequences of levels with the same quantum numbers.
This distribution reveals whether the system exhibits the so-called \uvo{level repulsion}, a phenomenon indicating the absence of additional integrals of motions.
We use the following formulae \cite{rmt2}:
\begin{equation}
\begin{array}{ccccc}
P_0(s)=\exp(-s)
& \leftrightarrow &
P_\omega(s)={\cal N}_\omega\ s^\omega\exp\left(-\alpha_\omega s^{\omega+1}\right)
& \leftrightarrow &
P_1(s)=\tfrac{\pi}{2}\ s\ \exp\left(-\tfrac{\pi}{4}s^2\right)\,,
\\ \vspace{-3mm} &&&& \\
{\rm Poisson} & & {\rm Brody} & & {\rm Wigner}
\end{array}
\label{brod}
\end{equation}
where the Brody form with $\omega\in(0,1)$ interpolates between the limiting cases $\omega=0$ (Poisson) and $\omega=1$ (Wigner).
Note that the symbol ${\cal N}_\omega$ stands for the normalization factor and $\alpha_\omega$ for a coefficient ensuring that $\ave{s}=1$.
The Poissonian distribution, which describes an uncorrelated spectrum, is typical for integrable systems.
It implies the absence of level repulsion (since $P_0\to 1$ for $s\to 0$), which is related to the existence of additional integrals of motions that eliminate the corresponding offdiagonal matrix elements of the Hamiltonian.
The Wigner formula includes the level repulsion given by $P_1\propto s$ for small spacings and captures properties of chaotic systems (with generally nonvanishing offdiagonal Hamiltonian matrix elements).\footnote
{
The Wigner formula in Eq.\,(\ref{brod}) is only an approximation of the distribution predicted by the random matrix theory. It applies in system symmetric under the time reversal. For other systems, the formula is modified to include even stronger ($\propto s^2$) level repulsion \cite{rmt2}.  
}
System in between order and chaos are characterized by a transitional Brody form with the parameter $\omega$ somehow related to the overall degree of chaos.
To be more specific, we expect a correlation (generally nonlinear) between values $(1-\omega)\ \leftrightarrow\ f_{\rm reg}$. 

\begin{figure}
\includegraphics[height=0.25\textheight]{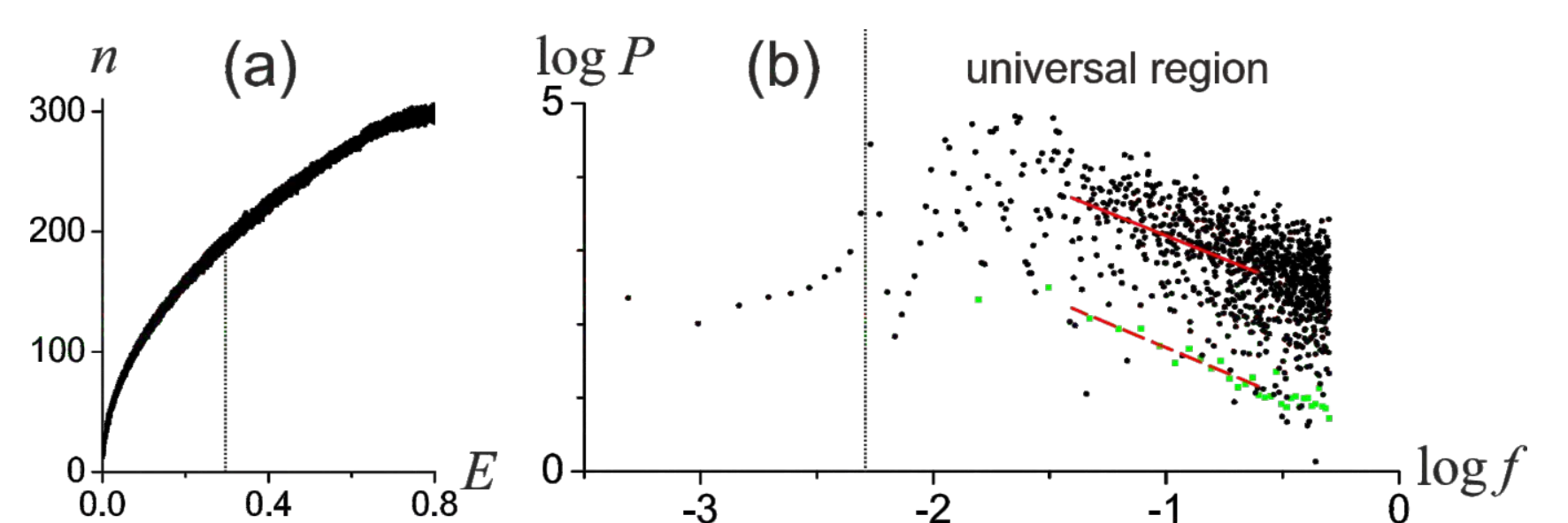}
\caption{Calculation of the power-spectrum exponent $\alpha(E)$ associated with the GCM unfolded energy spectrum at a single energy $E$. Model parameters: $(A,B,C)=(0.25,1,1)$, $\hbar^2/2M=5\cdot 10^{-7}$. Panel (a) displays the number $n$ of levels in the energy interval $\Delta E=2\pi\hbar/T$ corresponding to the shortest periodic orbit (period $T$) as a function of energy $E$. The value $n(E)$ sets the minimal frequency $f_{\rm min}(E)$ for the evaluation of the power spectrum. Panel (b) shows in the log-log scale the power spectrum $P_f$ for a sample of 2048 levels in a vicinity of the energy indicated in panel (a) by the vertical line. The slope of the cloud of points in the \uvo{universal region} above $f_{\rm min}$ (vertical line in panel~b) yields the exponent in the $\ave{P_f}\approx 1/f^{\alpha}$ dependence. The power spectrum is determined by two different methods: (i) by considering all 2048 levels (black dots) and (ii) by grouping the levels into 32 bunches of 64 levels and calculating an average power spectrum over the bunches (green dots). The corresponding fits of the slope, indicated by the full (i) and dashed (ii) tilted lines, yield values $\alpha^{\rm (i)}=1.27$ and $\alpha^{\rm (ii)}=1.32$. Note that in Fig.\,\ref{Qchagcm} we used the method (ii).}
\label{onef}
\end{figure}

Long range correlations in spectra of chaotic systems are also very strong.
They can be expressed in different ways, for instance via the variance of the number of unfolded levels in an interval of length $L$, or equivalently via a so-called $\Delta_3$ statistics \cite{rmt2}.
An alternative approach \cite{noise} is based on the Fourier analysis of a sequence of statistical quantities $\delta_n=e_n-e_1-n$.
The sequence apparently satisfies $\ave{\delta_n}=0$, expressing fluctuations of the cumulative number $N(e)$ of unfolded levels around the straight line $\bar{N}=e$.
The Fourier transformed sequence reads as $\tilde{\delta}_f=M^{-1}\sum_n\delta_n\exp[-2i\pi f n/M]$, with $M$ being the number of levels included into the analysis, and yields a power spectrum $P_f=|\tilde{\delta}_f|^2$.
The power spectrum contains information on the frequencies $f$ present in the \uvo{signal} $\delta_n$ (number $n$ being identified with a discrete time) and represents a statistical quantity associated with $\delta_n$.
It turns out that the average of the power spectrum can be well approximated by the formula $\ave{P_f}\approx 1/f^{\alpha}$, where $\alpha=1$ for strongly correlated spectra of chaotic systems, and $\alpha=2$ for the uncorrelated Poissonian spectra of regular systems \cite{noise}.
Hence we establish a link $(\alpha-1)\ \leftrightarrow\ f_{\rm reg}$.

Statistical properties of spectra were extensively studied for both the geometric collective model and the interacting boson model \cite{Str09a,Str09b,Mac07,Alh91,Alh93}.
Figure~\ref{Qchagcm} shows some of the GCM results, namely a comparison of classical regular fraction with the corresponding measures of short-range and long-range spectral correlations for two choices of the Hamiltonian parameters.
A clear correspondence between $f_{\rm reg}$ and the dependences of $(1-\omega)$ and $(\alpha-1)$ is observed, which confirms the above given statement concerning the link between chaos, on the classical level, and statistical properties of spectra, on the quantum level.
This \uvo{Bohigas conjecture}, as some authors call it, has been already tested in numerous systems, but mostly in the regime of a weak energy dependence of the regular fraction \cite{Boh84}.
In particular, the most influential analyses have been performed with 2D billiard systems, which show a constant value of $f_{\rm reg}$, totally independent of $E$.
Here we verify the Bohigas conjecture in a more complex form (though still based on a 2D model), demonstrating that even rapid changes of the classical regularity are closely followed by both short- and long-range quantum correlation measures.
Moreover, the comparison is performed for different quantization schemes and shows that the correlation properties are quantization-independent.
These properties must therefore reflect only the features related to the classical limit of the system, in agreement with the Bohigas conjecture.

We want to emphasize that the rapid variability of the GCM regular fraction with energy not only creates a desirable framework for testing the quantum chaos assumptions, but also represents a considerable computational challenge!
The statistical analysis of the spectrum can only be performed piece by piece---on samples of levels situated in a close vicinity of energy $E$, where $E$ varies along the whole spectrum.
On the one hand, the samples have to be sufficiently large to allow for a reliable statistical evaluation, on the other hand, the energy interval $\Delta E$ spanned by each sample needs to be sufficiently narrow with respect to the scale that determines local variations of $f_{\rm reg}(E)$. 
As an example, we illustrate in Fig.\,\ref{onef} the Fourier analysis of a single sample of levels around one particular energy $E$, outlining the procedures capable to determine the local power-spectrum exponent $\alpha(E)$. 

In the rest of this section we describe an interesting visual approach to quantum chaos, which will complement the cute Poincar{\' e} maps presented above in connection with classical chaos.
The method was proposed by A.\,Peres \cite{Per84b} (also in 1984, which seems to be {\it Annus Mirabilis\/} for quantum chaos) and its practical application is limited to systems with two degrees of freedom.
It is based on the fact that the time average $\bar{P}$ of an arbitrary quantity $P$ represents an automatic integral of motion, regardless of any specific dynamical properties of the system.
Classically, the average is defined through infinite-time integration over the whole classical orbit, and therefore assigns a unique value $\bar{P}(q,p)$ to any point of the phase space (the same value is assigned to all points visited by a single trajectory).
The integrals of motions $\bar{P}$ do not generally satisfy the conditions for integrability (any system has an infinite number of such integrals), but, as shown below, they provide a valuable analyzing tool for systems between order and chaos. 

\begin{figure}
\includegraphics[height=0.3\textheight]{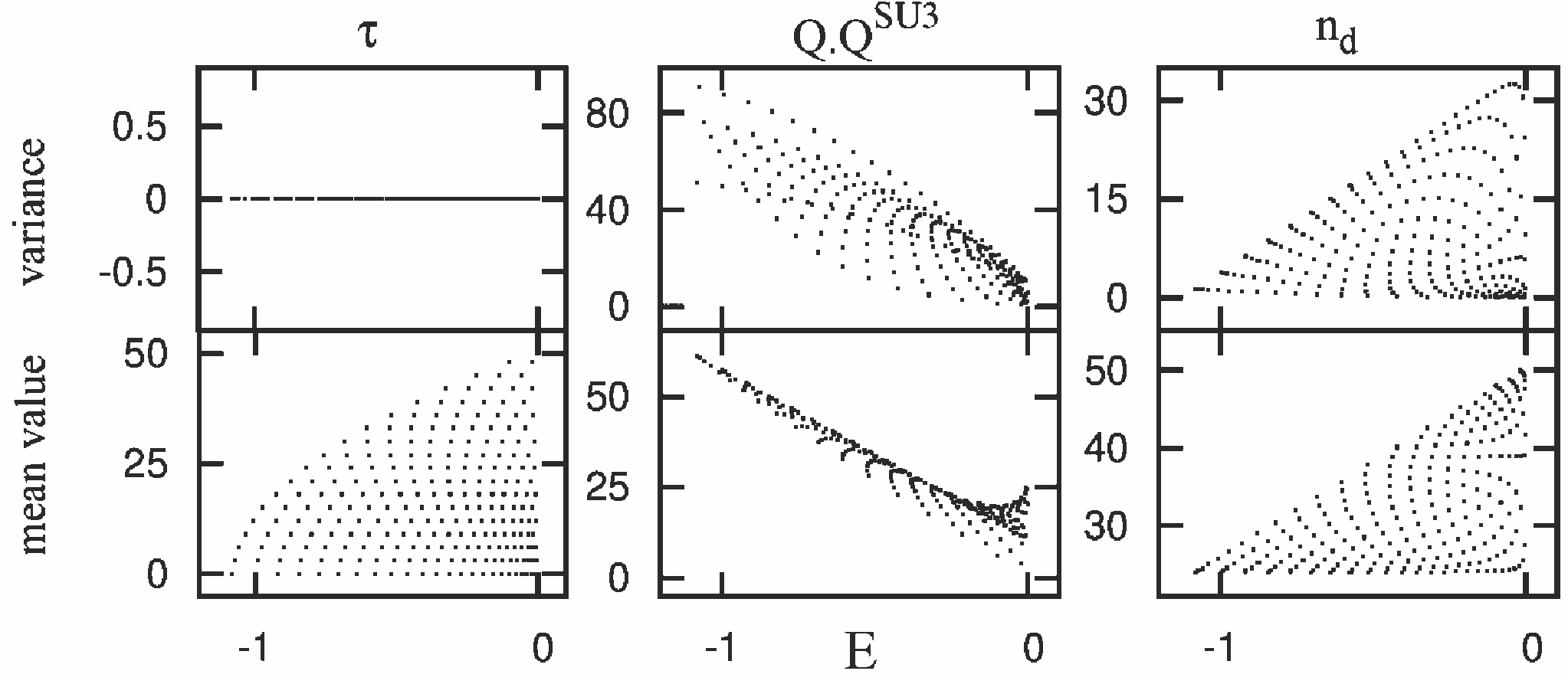}
\caption{Peres lattices of different quantities (indicated above each column) calculated for an integrable IBM Hamiltonian (\ref{Hibm}) with $A=0=\chi$ [the O(6) dynamical symmetry limit] for $N=50$ bosons. Symbol $\tau$ stands for the O(5) seniority, while the other  symbols denote Casimir invariants of the SU(3) and U(5) algebras, which are involved in the overall U(6) dynamical algebra of the IBM. The lower row shows the lattices $E_i\times \ave{P}_i$ of mean values of the respective quantities $P$, while the upper row shows the corresponding lattices of the variances $\ave{P^2}_i-\ave{P}^2_i$. 
%The variance of of $\tau$ is identically zero due to the compatibility of O(6) and O(5) invariants. 
Courtesy of M. Macek.}
\label{peribm}
\end{figure}

\begin{figure}
\includegraphics[height=0.65\textheight]{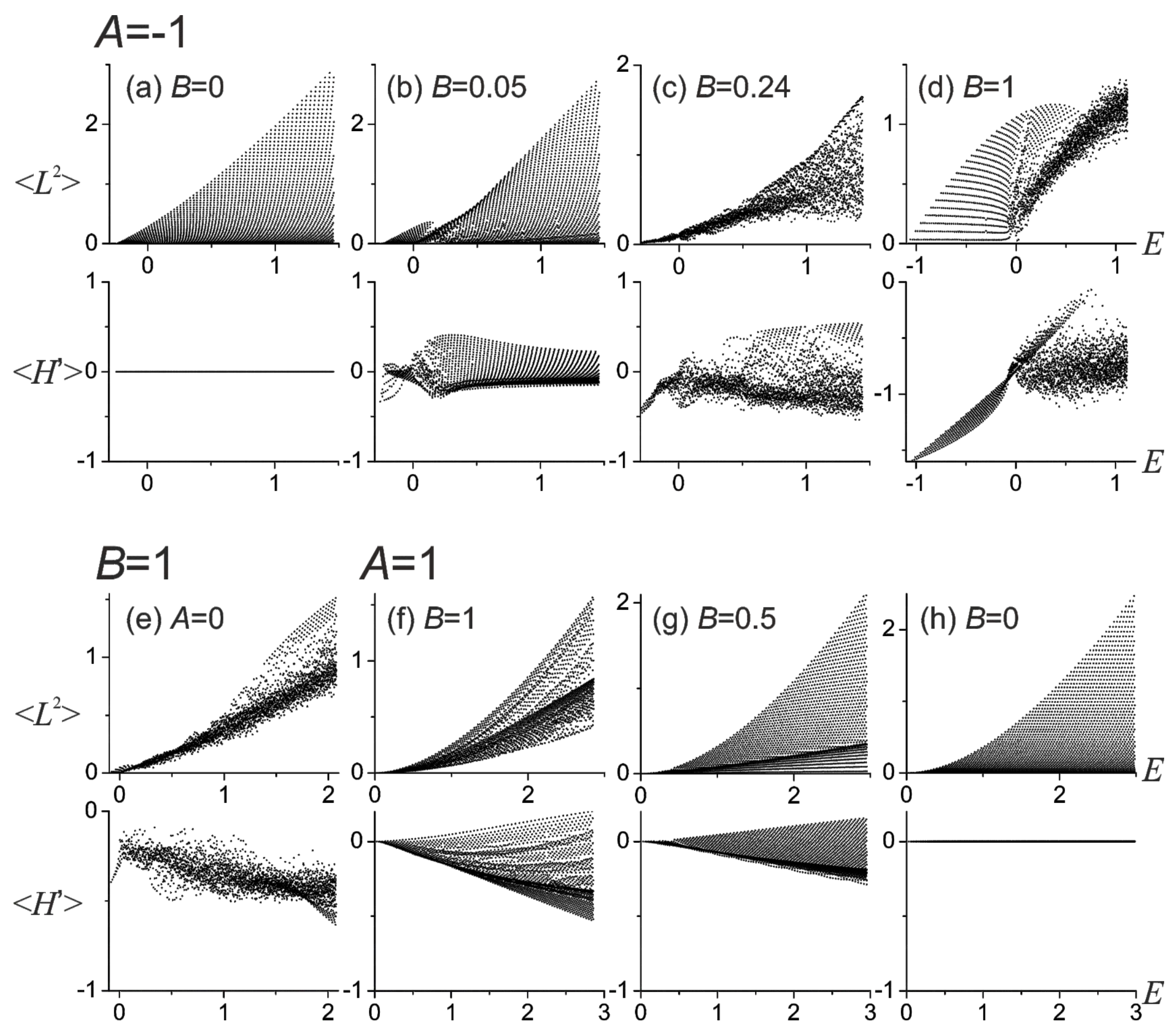}
\caption{Peres lattices of the GCM states with $J=0$ in the 5D quantization with $\hat{P}=\hat{p}_\gamma^2$ (upper rows) and $\hat{P}=\beta^3\cos 3\gamma$ (lower rows). Parameters $A$ and $B$ of the potential are indicated above the panels, $C=1$, and $\hbar^2/2M=3.2\cdot 10^{-5}$. }
\label{pergcm}
\end{figure}

In the quantum case, the time average is defined by the following formula:
\begin{equation}
\hat{\bar{P}}=\lim\limits_{T\to\infty}\frac{1}{T}\int_{0}^{T}dt\ \exp[+\tfrac{i}{\hbar}\hat{H}t]\,\hat{P}\,\exp[-\tfrac{i}{\hbar}\hat{H}t]
\,,\qquad
[\hat{\bar{P}},\hat{H}]=0
\label{per}
\,,
\end{equation}
where $\hat{P}$ is an operator corresponding to the original observable.
It is easy to verify that the commutator of the time-average operator $\hat{\bar{P}}$ with the Hamiltonian $\hat{H}$ vanishes in the $T\to\infty$ limit, irrespective of the concrete choice of $\hat{P}$.
Quantum expectation values of both observables $\hat{P}$ and $\hat{\bar{P}}$ are the same for the energy eigenstates of the system, $\ave{P}_i=\matr{E_i}{\hat{\bar{P}}}{E_i}=\matr{E_i}{\hat{P}}{E_i}$, and it is therefore rather straightforward to create a 2D picture in which the discrete energies $E_i$ are plotted against the corresponding expectation values $\ave{P}_i$.
For the resulting lattice of points we use the term \uvo{Peres lattice}.
It can be constructed for any observable $\hat{P}$.

Peres lattices for systems with $f=2$  can be seen as quantum counterparts of the classical Poincar{\' e} maps.
It turns out that geometric regularity (irregularity) of the lattice signals regular (chaotic) dynamics of the system on the classical level.  
In particular, if the system is classically integrable, any integral of motion must be a function (suppose a \uvo{well behaved} function) of the canonical actions, hence $H=H(I_1,I_2)$ and $\bar{P}=\bar{P}(I_1,I_2)$.
In transition to the quantum regime, $I_1$ and $I_2$ become approximate multiples of the Planck action and form a rectangular grid in the $I_1\times I_2$ plane.
As the Peres lattice in the $E\times\ave{P}$ plane is related to this $I_1\times I_2$ grid by a sooth mapping, it will certainly be somehow distorted, but one expects that it will still look orderly.
For non-integrable systems, on the other hand, there is no reason for the Peres lattice to be regular.\footnote
{
In the chaotic case, the expectation values of $\bar{P}$ for given energy $E$ tend to be contracted to a single value due to ergodicity of the system. 
}
In this case, any partial arrangement of the lattice is a signature of fractional regularity of the dynamics.
It should be noted that various Peres lattices can in general be combined with each other or smoothly transformed without affecting regularity or chaoticity of the patterns present in the lattice.
For instance, the lattice of $E_i$ versus $\ave{P^2}_i-\ave{P}^2_i$, which displays the variance of the quantity $P$ in the Hamiltonian eigenstates, is just a combination of lattices for $P^2$ and $P$.

Some examples of Peres lattices for the above IBM and GCM Hamiltonians are given in Figs.\,\ref{peribm} and \ref{pergcm}.
The first of them, Fig.\,\ref{peribm}, shows several incompatible lattices calculated for an integrable IBM Hamiltonian with the O(6) dynamical symmetry.
The O(5) seniority, cf. Eq.\,(\ref{senior}), whose average and variance lattices are drawn in the first column, represents a conserved quantity for any IBM Hamiltonian (\ref{Hibm}) with $\chi=0$, as can be directly verified through the null variance of the seniority in all Hamiltonian eigenstates.
The other quantities are not integrals of motions, they nevertheless generate perfectly regular lattices in both average values and dispersions.
This is in agreement with integrability of the Hamiltonian employed.

Figure~\ref{pergcm} depicts several Peres lattices for the GCM, using eigenstates with $J=0$  in the 5D quantization.
The lattices were constructed using only the average values of the respective Peres operator, the corresponding variances are not given this time.
Two choices of the Peres operator are presented:
The first one is the $J=0$ seniority 
%$\hat{P}=-\frac{\hbar^2}{\sin 3\gamma}\frac{\partial}{\partial\gamma}\sin 3\gamma\frac{\partial}{\partial\gamma}$, 
$\hat{P}=-\hbar^2\sin^{-1}\!3\gamma\,(\partial/\partial\gamma)\sin 3\gamma\,(\partial/\partial\gamma)$, 
which expresses an \uvo{angular momentum} operator $\hat{p}^2_\gamma$ in the plane $x\times y$ of the deformation coordinates (in the 5D quantization). 
The corresponding averages are denoted as $\ave{L^2}$ and their lattices are shown in the first row of each panel.
The second choice of the Peres operator coincides with the cubic term of the GCM potential, $\hat{P}=\beta^3\cos 3\gamma$, which represents the nonintegrable perturbation of the $B=0$ integrable Hamiltonian. 
The averages, denoted $\ave{H'}$, are seen in the second row of each panel.

Peres lattices in Fig.\,\ref{pergcm} are given for several points in the parameter plane $A\times B$, following a path which crosses all inequivalent GCM potentials differing by the scale invariant parameter $\tau$.
At first, we fix $A=-1$ and change $B=0\to 1$, visualizing the onset of chaotic vibrations in deformed nuclei and their partial regularization at larger values of $B$. 
Next we keep $B=1$ and change $A=-1\to 0\to+1$. 
In this way, we make a transition between two qualitatively different types of the potential---the one corresponding to deformed equilibrium shapes of the nucleus and the one associated with spherical nuclei
(the $A<0$ potentials have three degenerate minima at $\beta_0>0$, while those with $A$ above a certain critical value $A_c>0$ have just a single global minimum at $\beta_0=0$).\footnote
{
The \uvo{critical point} where the $\beta_0=0$ and $\beta_0>0$ minima of the potential become degenerate and swap is at $(A,B)=(0.25,1)$.
}
Finally, we fix $A=+1$ and decrease $B=1\to 0$, coming back to the integrable regime, but on the spherical side of the parameter plane.

The most chaotic lattices are the ones in panels (c) and (e).
The first corresponds to $(A,B)=(-1,0.24)$, which coincides with the first minimum of $f_{\rm reg}$ at $E=0$ in Fig.\,\ref{freg0}.
Both Peres lattices in this panel can be compared with the energy dependences of $f_{\rm reg}(E)$ and with the other measures of chaos shown in panels (a) and (c) of Fig.\,\ref{Qchagcm}.
We may notice that the regions with increased regularity in Fig.\,\ref{Qchagcm} correspond to the occurrence of regular parts of the lattice in Fig.\,\ref{pergcm}(c).
The chaotic lattice in panel (e) corresponds to $(A,B)=(0,1)$, which is a point situated near the \uvo{phase transition} between spherical and deformed shapes.
Many other examples of Peres lattices and their detailed comparison with the dependence $f_{\rm reg}(E)$ can be found in Refs.\,\cite{PStr} and \cite{Str09b}.

From the examples shown here and in the literature it can be inferred that  Peres lattices significantly complement the standard methods used to investigate chaotic properties of quantum systems.
In fact, they even enable one to go beyond the standard methods.
This is so because quantum states can be \emph{individually assigned} to distinctly regular or distinctly chaotic parts of the lattice, which is in contrast to the statistical approach that can only attribute some averaged regular/chaotic attributes to sufficiently large ensembles of states.
Constructing a suitable Peres lattice for a given quantum system with known spectral correlations is therefore similar to analyzing the classical dynamics with a given $f_{\rm reg}$ by means of Poincar{\' e} sections.

Two remarks on the applicability of the method are in order:
First, to unambiguously deduce whether a given state is in a regular or chaotic domain of the spectrum, it is usually useful to construct Peres lattices for several operators.
A comparison of several lattices helps particularly in situations when an overlap of two or more regular parts in a single lattice produces a seemingly chaotic mesh.
Second, the method  in the above-presented form is applicable only in systems with $f=2$.
Consider, for example, an integrable system with $f=3$. 
To ensure a one-to-one mapping from the space of canonical actions $(I_1,I_2,I_3)$, one would have to consider a $3$-dimensional lattice ($f$-dimensional in general).
This is, however, rather difficult with only 2D physical literature being available.
These unalterable limitations of us, poor 3D creatures, will probably always hinder our understanding of higher than 2D systems.

\section{Conclusions}

We are arriving to the end of our short tour.
Its purpose was to show some known concepts of chaos theory by means of simple models taken from the nuclear context.
Was the marriage of chaos with nuclei successful?
We believe it was.
We think that the models like those used here provide an excellent environment for general investigations of chaos in both classical and quantum incarnations---environment, which is sufficiently complex to embrace diverse nontrivial phenomena, but at the same time simple enough to allow an intelligible description.
In analogy with the billiard models, that have been extensively used for such investigations in the past decades, the nuclear models can be used as models with two degrees of freedom.
In contrast to the billiard models, however, they exhibit very rich dependence of chaos measures on both energy and Hamiltonian parameters, which opens some new directions of potential research.

We tried to make the presentation comprehensible, but it could not be comprehensive.
Among the items that we left intact here, the following are particularly worth mentioning:
\begin{itemize}
\item Geometrical chaos: 
Attempts have been made to develop techniques that would be able to \emph{predict} an approximate value of the regular fraction $f_{\rm reg}$ for an arbitrary choice of the system's parameters and energy. 
An interesting approach is based on the geometrical method, which reduces the dynamics of the system to free motions of particles in curved space \cite{geom1}. 
A related technique resorts to evaluating the convexity/concavity of the energy contours of the model potential \cite{Str06}. 
It can be shown that in the GCM (and/or IBM) the geometrical method is only partially successful \cite{geom2}.  
Pre-estimation of the degree of chaos in a given system therefore remains one of the major tasks of chaos theory.
\item Chaos versus symmetry: 
People keep trying to establish a link of order and chaos with some generalized concepts of symmetry. 
If such a link exists and is universal, it would interpret partial regularity, as observed in various systems of nature, through some imperfect, fractional dynamical symmetries possessed by these systems. 
This led to the ideas of so-called partial dynamical symmetry \cite{pds} and quasi dynamical symmetry \cite{qds}. 
Although both GCM and IBM are particularly well suited for this kind of research \cite{qds2,qds3}, its success in connection with chaos is so far only partial. 
\item Implications of chaos: 
One can step from origins to consequences of chaos, investigating various chaos-assisted phenomena in specific systems. 
For example, the coexistence of regular and chaotic modes of dynamics turned out to be relevant for transitions between spherical and deformed shapes of nuclei \cite{bolo,mami}. 
Regular dynamics was also shown to play a remarkable role in the emergence of rotational bands in highly excited nuclei \cite{sepa}, which suggests that the physics underlying the general phenomenon of adiabatic separation of intrinsic and collective motions in many-body systems might be closely related to chaos theory. 
\end{itemize}

%%%%%%%%%%%%%%%%%%%%%%%%%%%%%%%%%%%%%%%%%%%%%%%%
%% BACKMATTER
%%%%%%%%%%%%%%%%%%%%%%%%%%%%%%%%%%%%%%%%%%%%%%%%

\begin{theacknowledgments}
This work was supported by the Czech Science Foundation (project no.\,P203-13-07117S), as well as by CONACyT, Mexico (project no.\,155663) and PAPIIT-UNAM, Mexico (project no.\,IN114411).
\end{theacknowledgments}



\bibliographystyle{aipproc}   % if natbib is available


%\bibliography{sample}

 %{\typeout{}
  %\typeout{******************************************}
  %\typeout{** Please run "bibtex \jobname" to optain}
  %\typeout{** the bibliography and then re-run LaTeX}
  %\typeout{** twice to fix the references!}
  %\typeout{******************************************}
  %\typeout{}
 %}

\end{document}